\documentclass[%
 reprint,
 amsmath,
 amssymb,
 aip,
 noeprint,
 floatfix
]{revtex4-2}
\usepackage[utf8]{inputenc}
\usepackage{graphicx}
\usepackage{subfigure}
\usepackage{xcolor}
\usepackage{amsmath}
\usepackage{amssymb}
\usepackage{hyperref}
\hypersetup{colorlinks=true, allcolors=blue}
\usepackage{siunitx}
\usepackage{import}
\usepackage{newtxtext,newtxmath}
\usepackage{microtype}

\newcommand{\partFig}[2]{Fig.~\hyperref[#1]{\ref*{#1}(#2)}}
\newcommand{\eg}{\textit{e.g.}}
\newcommand{\ie}{\textit{i.e.}}
\newcommand{\loss}{G^{s\prime\prime}}
\newcommand{\elastic}{G^{s\prime}}
\newcommand{\pnip}{PNIPAM-SiO$_2$}
\newcommand{\mpc}{{\rm mol.\,\%}}
\newcommand{\wpc}{{\rm wt}\,\%}
\newcommand{\vpc}{{\rm vol.}\,\%}

\begin{document}

\title{Contactless Interfacial Rheology: Probing Shear at Liquid-Liquid Interfaces without an Interfacial Geometry via Fluorescence Microscopy }

\author{Iain Muntz}
\thanks{These authors contributed equally to this work}
    \affiliation{%
    SUPA School of Physics and Astronomy,\\
    The University of Edinburgh, Edinburgh,\\
    EH9 3FD, Scotland, United Kingdom
    }%
    \affiliation{%
    Department of Bionanoscience, Kavli Institute of Nanoscience Delft, \\
    Delft University of Technology, Van der Maasweg 9, \\
    2629 HZ Delft, The Netherlands
    }%

\author{James A. Richards}
\thanks{These authors contributed equally to this work}
    \affiliation{%
    SUPA School of Physics and Astronomy,\\
    The University of Edinburgh, Edinburgh,\\
    EH9 3FD, Scotland, United Kingdom
    }%
    \affiliation{Edinburgh Complex Fluids Partnership, The University of Edinburgh,\\
    Edinburgh EH9 3FD, United Kingdom}

\author{Sam Brown}
    \affiliation{%
    SUPA School of Physics and Astronomy,\\
    The University of Edinburgh, Edinburgh,\\
    EH9 3FD, Scotland, United Kingdom
    }%

\author{Andrew B. Schofield}
    \affiliation{%
    SUPA School of Physics and Astronomy,\\
    The University of Edinburgh, Edinburgh,\\
    EH9 3FD, Scotland, United Kingdom
    }%

\author{Marcel Rey}
    \affiliation{%
    SUPA School of Physics and Astronomy,\\
    The University of Edinburgh, Edinburgh,\\
    EH9 3FD, Scotland, United Kingdom
    }%
    \affiliation{%
    Department of Physics, \\
     University of Gothenburg, Gothenburg, Sweden\\}

\author{Job~H.~J.~Thijssen}
    \email{j.h.j.thijssen@ed.ac.uk}
    \affiliation{%
    SUPA School of Physics and Astronomy,\\
    The University of Edinburgh, Edinburgh,\\
    EH9 3FD, Scotland, United Kingdom
    }%

\begin{abstract}
    Interfacial rheology is important for understanding properties such as Pickering emulsion or foam stability. Currently, the response is measured using a probe directly attached to the interface. This can both disturb the interface and is coupled to flow in the bulk phase, limiting its sensitivity. We have developed a contactless interfacial method to perform interfacial shear rheology on liquid/liquid interfaces with no tool attached directly to the interface. This is achieved by shearing one of the liquid phases and measuring the interfacial response via confocal microscopy. Using this method we have measured steady shear material parameters such as interfacial elastic moduli for interfaces with solid-like behaviour and interfacial viscosities for fluid-like interfaces. The accuracy of this method has been verified relative to a double-wall ring geometry. Moreover, using our contactless method we are able to measure lower interfacial viscosities than those that have previously been reported using a double-wall ring geometry. A further advantage is the simultaneous combination of macroscopic rheological analysis with microscopic structural analysis. Our analysis directly visualizes how the interfacial response is strongly correlated to the particle surface coverage and their interfacial assembly. Furthermore, we capture the evolution and irreversible changes in the particle assembly that correspond with the rheological response to steady shear.
\end{abstract}

\maketitle

\section{Introduction}
Interfacial rheometry is essential when characterising systems with large interfacial area, such as emulsions or foams~\cite{Fuller2012,Thijssen2018,Derkach2009}. These systems are ubiquitous in industries such as pharmaceuticals, cosmetics and foodstuffs~\cite{Binks2006, Tavacoli2015, Leal-Calderon2008, Hunter2008,Tcholakova2008, fernandez2021microgels, Perrin2020}. In order to probe the rheological properties, one can use shear rheology~\cite{Vandebril2010, Masschaele2011, Keim2013, VanHooghten2017,Kragel2010}, dilational rheology~\cite{Brugger2010,Razavi2015,Garbin2015,Ravera2010}, or simultaneously image the interface as shear is applied to connect the rheological properties to the interfacial microstructure \cite{Barman2014, Barman2016}.

Previous work on interfacial shear rheology has used probes which directly attach to an oil-water or air-water interface, such as the magnetic rod interfacial stress rheometer \cite{Brooks1999, Reynaert2008,Tajuelo2015}, or the double-wall ring (DWR) geometry attached to a rotational rheometer \cite{Vandebril2010, Barman2014}. These experimental setups are both based on the maximisation of the ratio of the surface force to the sub-phase drag, the Boussinesq number \cite{Fuller2012, Jaensson2021}:
\begin{equation}
    \label{eq:boussinesq}
    \mathrm{Bo} = \frac{\eta^s}{\eta l},
\end{equation}
where $\eta^s$ is the surface viscosity, $\eta$ is the sub-phase viscosity and $l$ is a characteristic length scale roughly equal to the ratio of contact area to contact perimeter. In order to accurately measure the surface properties without unintentionally probing the sub-phase, this ratio must be maximised for the surface to contribute at least an order of magnitude more than the bulk. Of the two setups mentioned, the magnetic rod has the larger Bo, while both have a Bo an order of magnitude larger than that of a rotating disk, due to a much smaller $l$~\cite{Kim2011}. This maximisation of Bo can be considered as optimising the interface-to-bulk signal to noise ratio. Even though the magnetic rod set up has higher sensitivity, DWR has the advantages of using a conventional rotational rheometer combined with a larger dynamic range~\cite{Vandebril2010}.

In our work, we take a different approach which makes consideration of Bo less tangible as we have no contact area or contact perimeter. Rather than affixing a probe directly to the interface, we shear the upper phase, indirectly deforming the interface, and measure the response using confocal microscopy: a fundamentally different approach.

Using this contactless technique, we investigate the efficacy of this method by studying a jammed core-shell \pnip--laden interface labelled with tracer particles, which we compare directly to DWR measurements. We then demonstrate the advantages of this technique by looking at a weakly interacting system of interfacially adsorbed colloidal particles. This system has been studied previously using direct probe techniques~\cite{VanHooghten2017,Mears2020}, and considering the interparticle interactions~\cite{Muntz2020}.

Our technique has two main advantages: (i) the liquid-liquid interface we probe is not disturbed by a large probe immersed therein, and (ii) this setup models general applications of these large interfacial area systems, where interfacial shear is applied indirectly via the continuous phase. A clear example of this second point is in the application of skin creams, where the continuous phase is sheared, which indirectly deforms the large area of interface of the dispersed phase. Notably, the equipment required to perform these measurements is relatively common. While we use confocal microscopy coupled to a stress-controlled rheometer, reflection or fluorescence microscopy and a fixed-rate motor should suffice. We show that our technique can measure surfaces with lower viscosities than have been measured before using a DWR geometry, due to the inherent sensitivity of the technique arising from the absence of direct sub-phase drag. Finally, our setup lends itself to simultaneous structural analysis, which we show is key to understanding the rheological properties of a particle-laden interface.

\section{Materials and Methods}\label{sec:mat_and_methods}
\subsection{Materials}

All chemicals were obtained from commercial sources and used as received if not stated otherwise. N,N’-Methylenebis(acrylamide) (BIS; 99 $\%$, Sigma Aldrich), ethanol (EtOH, 99.9 $\%$, Sigma Aldrich), ammonium persulfate (APS; 98 $\%$ Sigma Aldrich), tetraethyl orthosilicate (TEOS; 98 $\%$, Sigma Aldrich), ammonium hydroxide solution (28-30 $\%$ NH$_{3}$ basis, Sigma Aldrich), (3-(trimethoxysilyl)propyl methacrylate (MPS; 98 $\%$, Sigma Aldrich) and isopropyl alcohol (IPA, $>99.8$ $\%$, Sigma Aldrich), were used as received. 

N-Isopropylacrylamide (NIPAM; 97 $\%$, Sigma Aldrich) was purified by recrystallization from hexane (95 $\%$, Sigma Aldrich). 
Water was distilled and deionized (\SI{18} {\mega\ohm\centi\metre}) and \textit{n}-dodecane (Acros organics, 99\% pure) was filtered three times through a column of alumina (Sigma-Aldrich, activated) to remove polar impurities following a standard procedure~\cite{Goebel1997}. Red fluorescent carboxyl-functionalized polystyrene (PS) particles (\SI{2}{\micro\metre} diameter, Thermo Fisher) were cleaned twice via centrifugation and redispersion in water/ethanol (1:1).

\subsection{Synthesis and Characterisation}
\subsubsection{\texorpdfstring{PNIPAM-SiO$_{2}$ core-shell particles}{PNIPAM-SiO2 core-shell particles}} 
Poly(N-isopropylacrylamide)(PNIPAM)-SiO$_{2}$ core-shell particles  were  obtained by growing a PNIPAM shell onto the silica cores via a batch surfactant-free precipitation polymerization as described in previous work~\cite{Ciarella2021}.  
First, colloidal silica particles used as cores with a diameter of \SI{160(10)}{\nano\metre} were prepared according to a modified St\"ober process\cite{Sing2018}. In a round bottom flask, 250\,mL EtOH , 12.5\,mL deionised water and 25\,mL NH$_{3}$ (aq) were stirred together. 18.75\,mL of TEOS was stirred in 75\,mL EtOH and both solutions were heated to $\SI{50}{\celsius}$ and equilibrated for \SI{30}{\minute}. Next, the TEOS solution was quickly added to the first mixture under heavy stirring. We let the reaction proceed for \SI{2}{\day} at $\SI{50}{\celsius}$. 
The suspension was functionalised without any further purification by adding \SI{102.7}{\micro\litre} MPS. We allowed the reaction mixture to stir at room temperature for at least \SI{1}{\day} and then boiled it for \SI{1}{\hour} to ensure successful functionalisation. Afterwards, we purified the particles by centrifugation and redispersed them three times in ethanol and three times in Milli-Q water. 

In a 500\,mL three-neck round bottom flask, 282.9\,mg NIPAM and 19.3\,mg BIS ($5\,\mpc$) were dissolved in 47\,mL Milli-Q water. We added the 2.591\,g aqueous SiO$_2$ core dispersion (6.6\,$\wpc$). The solution was heated to $\SI{80}{\celsius}$ and purged with nitrogen. After equilibration for \SI{30}{\minute}, a balloon filled with nitrogen was used to keep the nitrogen atmosphere. Subsequently, 11\,mg APS was rapidly added to initiate the reaction. We let the reaction proceed for 4\,h, and after it cooled down, we purified the suspension 6$\times$ by centrifugation and redispersion in deionised water. 
The hydrodynamic diameter at \SI{20}{\celsius} was determined by dynamic light scattering (Malvern Zetasizer Nano-ZS) to \SI{525(53)}{\nano\metre}.

\subsubsection{PMMA particles} 

Poly(methyl methacrylate) (PMMA) particles, stabilized by poly(lauryl methacrylate), were used as the hydrophobic system. To synthesize these, the poly(lauryl methacrylate) stabilizer was fabricated first following the recipe in Ref.~\onlinecite{Barrett1974}, Sec.~3.9.1, and it was kept as a 40\% solution in dodecane. To make the particles, a mixture was created that contained 2.1\%\,w/w poly(lauryl methacrylate) stabilizer, 41.2\% w/w methyl methacrylate, 0.84\% methacrylic acid, 11\% butyl acetate, 29.6\% hexane, 14.2\% dodecane, 0.21\% octyl mercaptan and 0.47\% of the dye NBD-MAA (7-nitrobenzo-2-oxa-1,3-diazole-methyl methacrylate), whose preparation can be found in Ref.~\onlinecite{Jardine2002}. This mixture was placed in a 3-necked round-bottomed flask with a condenser attached, brought under a nitrogen atmosphere, stirred at 350 rpm and heated to 80°C before 0.4\%\,w/w of the initiator azo-bis-isobutyronitrile was added to start the polymerization reaction which was left to proceed for 6 hours. The resultant particles were filtered through glass wool to remove any coagulum present. The particles were qualitatively inspected using scanning electron microscopy, and sized by static light scattering to find a diameter of \SI{3.0}{\micro\metre} with a dispersity of 5\%. The particles were cleaned by repeated centrifugation (5$\times$) in $n$-hexane followed by repeated centrifugation (5$\times$) in $n$-dodecane. The particles were kept as a dispersion in $n$-dodecane and sonicated for 30 minutes before dilution, followed by a further 30 minutes of sonication before use to minimise the number of aggregates in bulk.

\subsection{Contactless methods}

\begin{figure}
    \centering
        \includegraphics{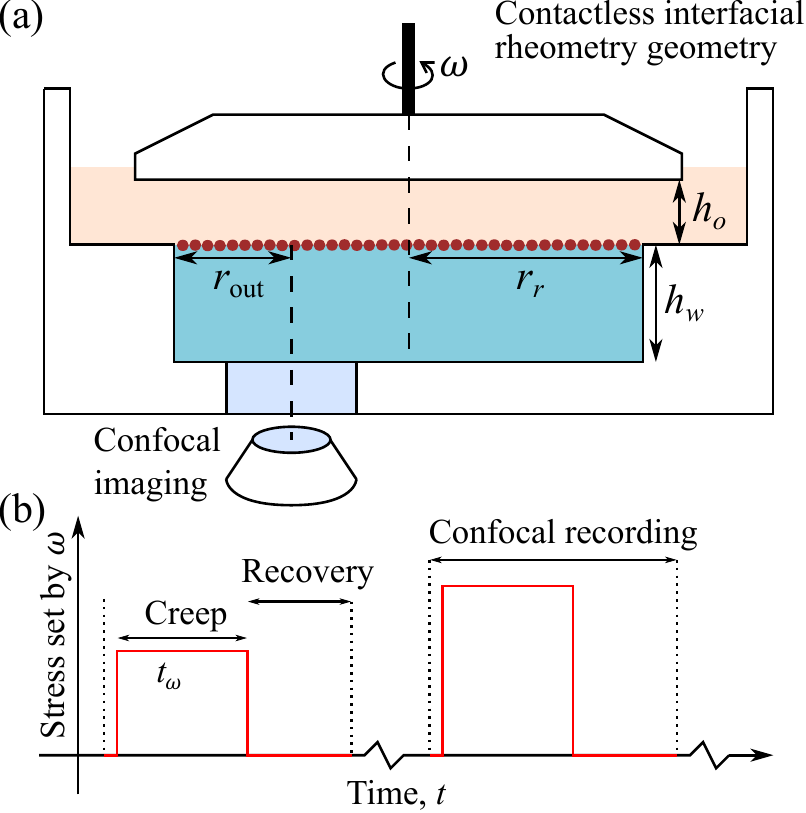}
        \includegraphics{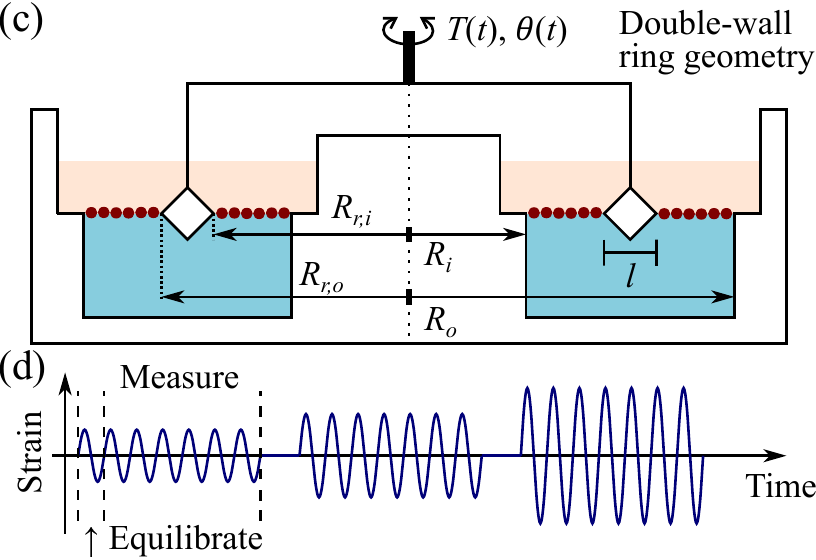}
    \caption{Interfacial shear geometries. (a)~Schematic setup for contactless interfacial rheology using a parallel-plate geometry rotated at fixed angular velocity, $\omega$. Interface imaging via confocal microscope. Dimensions: pinned interface ring radius, $r_r = 10$\,mm; water sub-phase height, $h_w = 3$\,mm; and oil depth, $h_o$. $\omega$ and $h_o$ vary. $r_{\rm out}$ distance from imaging region to outer edge varies with experiment from $\SIrange{4}{6}{\milli\metre}$. (b)~Creep-recovery protocol for contactless method. Recording (dashed lines) begins before applied $\omega$ (solid line) to $t_{\omega} = \SIrange{30}{60}{\second}$, followed by recovery to recording end. Multiple steps with increasing $\omega$. (c)~Schematic of double-wall ring geometry. Radii: ring, inner and outer, $R_{r,i} = \SI{35.5}{\milli\metre}$ and $R_{r,o} = \SI{34.5}{\milli\metre}$; trough, $R_{i} = \SI{31}{\milli\metre}$ and $R_o = \SI{39.5}{\milli\metre}$. Ring width, $l = R_{r,o}-R_{r,i} = \SI{1}{\milli\metre}$ for Bo, Eq.~\eqref{eq:boussinesq}. Torque, $T(t)$, and angle, $\theta(t)$, to calculate interfacial stress, $\sigma^s$, and strain, $\gamma^s_0$. (d)~Increasing logarithmic oscillatory strain amplitude sweep.}
    \label{fig:geometry}
\end{figure}

Interfaces were prepared in a custom made polytetrafluoroethylene (PTFE) cup with an aluminium ring insert to pin a flat interface. \partFig{fig:geometry}{a} shows a schematic representation of the cup. The PTFE cup's inner radius ($r_c$) is \SI{21}{mm}, the aluminium ring's inner radius ($r_r$) is \SI{10}{mm}. The aluminium ring has a height of \SI{3}{mm} and its edge was roughened using silicon carbide sandpaper to allow pinning of the interface.

The PTFE cup was filled with water, pinned at the edge of the aluminium ring. As the first interface, we choose a monolayer of PNIPAM-SiO$_2$ core-shell particles mixed with fluorescent tracer particles at a fixed surface pressure of \SI{24}{\milli\newton\per\metre}. We first created a suspension by mixing \SI{800}{\micro\litre} core-shell particles (0.1\,$\wpc$) and \SI{100}{\micro\litre} fluorescent PS microspheres (1\,$\wpc$) with \SI{100}{\micro\litre}~IPA as a spreading agent. The \pnip\ core-shell particles are smaller and able to spread and extend once adsorbed to liquid interfaces~\cite{Rey2020,Rauh2017}. Therefore, they occupy more interfacial area compared to the PS particles. Further, we should point out that PNIPAM-based microgels are known to adsorb onto PS particles \cite{Rey2020aggr} and accumulate around them when confined at liquid interfaces \cite{Rey2017jacs}. Thus, we expect the PS particles to be integrated within the \pnip\ interface and only minimally influence the rheological response.   We then spread the mixed suspension on a Langmuir trough and measure the surface pressure using the Wilhelmy method. We determined that \SI{3.72}{\micro\litre\per\centi\metre^2} of prepared suspension is required to obtain a surface pressure of \SI{24}{\milli\newton\per\metre}. Thus, \SI{11.7}{\micro\litre} was pipetted onto the air-water interface pinned by the aluminum ring. Notably, in a control experiment, we could directly verify the surface pressure of the interface within the cell using the Wilhelmy method to be \SI{24}{\milli\newton\per\metre}. Lastly, \SI{1.5}{\milli\litre} of dodecane was carefully pipetted on top of the interface to give a depth of the oil phase ($h_o$) of \SI{1.6}{\milli\metre}.

As a second interface, we choose hydrophobic PMMA colloidal particles. The aluminum ring was again filled with water and \SI{3}{\milli\litre} of a dilute 0.005\,$\vpc$ PMMA-in-oil dispersion was pipetted onto the water sub-phase ($h_o = \SI{2.2}{\milli\metre}$). A 0.005\,$\vpc$ dispersion generally leads to a low volume fraction interface, however there is large variability in the final surface fraction for the same initial volume fraction. To achieve higher surface fractions a higher initial volume fraction was used. During equilibration for 1 hour, the PMMA particles sedimented to the oil-water interface and formed a monolayer. The surface fraction, $\phi$, was adjusted by either adjusting the amount of deposited oil dispersion or by adjusting the particle concentration.

Surface fractions were measured from microscopy images using a pixel counting method determining the fraction of foreground (particle) pixels to total pixels after performing a thresholding procedure. This measurement was made over multiple frames and the final value for surface fraction was determined as the mean through one rotation of the interface. We note that this approach likely overestimates the actual $\phi$, see Appendix~\ref{app:surfaceFraction}. However, this simple approach allows systematic comparison between different surface coverages.

Connected pixel clusters can then be identified to assess interface homogeneity, or aggregation state, via the dispersity~\cite{VanHooghten2017},
\begin{equation}
    \mathcal{D} = \frac{s}{\langle A\rangle}\,,
    \label{eq:pd}
\end{equation}
with $\langle A \rangle$ the average cluster size and $s$ the standard deviation. These dispersities were measured from relatively zoomed out images.

A 25~mm diameter parallel-plate geometry was attached to the oil-air interface in the centre of the PTFE cup. Fixed rotation speeds, $\omega$, were then applied (MCR 301, Anton Paar), shearing the upper oil phase.

Using a rheoimaging setup, as described by \citet{Besseling2009} (although our rheometer setup lies directly on top of the confocal, as described by \citet{Dutta2013}, providing greater stability), rheometry was conducted while the interface was simultaneously imaged using a Leica SP8 confocal microscope with a 10$\times$ / 0.3 NA air-immersion objective, at $1024\!\times\! 1024$\,px$^2$ (\pnip) or $512\!\times\! 512$\,px$^2$ (PMMA) ($\SI{932}{\micro\metre} \! \times \! \SI{932}{\micro\metre}$). The imaging setup was such that the motion of the interface under shear was horizontally oriented. Velocimetry of the confocal images was performed using C code written in house. This splits the images into 10 equally spaced horizontal bands, \ie\ the top 10\% of the image, the second 10\% of the image, etc. Each band is offset horizontally by a well defined distance. The Pearson correlation coefficient of this new offset band with the same band in the previous frame is calculated~\cite{Hermes2013}. The distance moved between that frame and the previous is then the horizontal offset which maximises this correlation, over time this gives the net displacement. Note that, as we use fluorescence microscopy, and only the particles are fluorescently labelled, the strain we measure is the strain of the (interfacial) colloidal particles in the field of view of the microscope.

The interfacial strain is calculated as $x/r_{\rm out}$ for each band and averaged, where $x$ is the measured displacement of the interface and $r_{\rm out}$ is the distance from the measurement to the outer, pinned wall; $r_{\rm out}$ varies for each experiment and is measured \textit{in situ}, it is always approximately \SI{6}{\milli\metre}. To probe the yielding and flow of the interfaces, a creep-recovery protocol is used~\cite{Nguyen1992}. Fixed rotation rates were set for $t_{\omega} = \SI{60}{\second}$ (PNIPAM-SiO$_2$) or \SI{120}{\second} (PMMA), applying stress to the interface, before a further period of fixing the rotation of the rheometer to 0, \SI{60}{\second} (PNIPAM-SiO$_2$) or \SI{30}{\second} (PMMA), allowing the interface to relax, \partFig{fig:geometry}{b}. The two steps allow separate measurement of the elastic response and the plastic, or irrecoverable, response. The (elastic) recoverable strain ($\gamma^s_{\rm rec}$), is given by the recoil from the peak strain at the end of the applied rotation to the end of the recovery step. The (plastic) irrecoverable shear rate, $\dot\gamma^s$, can be calculated from the total change in strain from the start of applied rotation to after recovery over the time of the applied shear. Alternatively, for faster relaxing interfaces (PMMA) $\dot\gamma^s$ can be calculated from the average shear rate over a \SI{45}{\second} window towards the end of the applied rotation. This sequence is repeated at multiple increasing rotation rates to determine the stress-dependent response of the interfaces. 

\subsection{Double-wall ring geometry}

Conventional interfacial shear rheometry was performed using a double-wall ring geometry, \partFig{fig:geometry}{c}, connected to a stress-controlled rotational rheometer (TA Instruments, DHR-2). This consists of a Platinum/Iridium ring (diamond cross-section with inner/outer radius $R_{r,i/o} = 34.5/\SI{35.5}{\milli\metre}$ and hence width $l = \SI{1}{\milli\metre}$) inside a ring-shaped polyoxymethylene trough (inner/outer radius $R_{i/o}= 31/\SI{39.5}{\milli\metre}$). All surfaces were cleaned multiple times with ethanol and deionised water. To form an interface the trough is first filled with the water sub-phase until level and pinned at the edges of the trough. \SI{70}{\micro\litre} of the mixed \pnip\ and PS tracer particle dispersion in a spreading solvent were then carefully pipetted onto the air-water interface. The ring is then lowered until pinned and level at the interface, before dodecane is pipetted on top to cover the ring. The ring is therefore in direct contact with both the interface and the sub-phase.
\begin{figure}
    \includegraphics{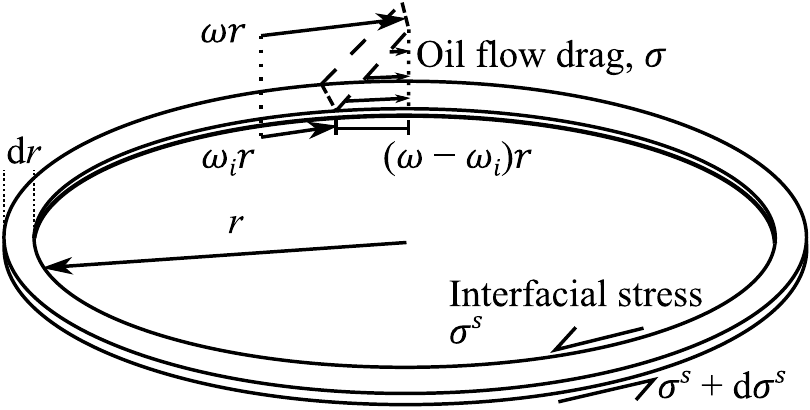}
    \caption{Thin ring element of oil-water interface, radius $r$ and width d$r$. The interfacial stress, $\sigma^s$, can be found by considering the torque balance of bulk oil flow drag and interfacial stress gradient, d$\sigma^s$, with rotation rates (interface, $\omega_i$, and top of oil phase, $\omega$), Sec.~\ref{sec:shear}.}
    \label{fig:torqueBalance}
\end{figure}
Oscillatory strain amplitude sweeps were performed using controlled strain at $f = \SI{0.2}{\hertz}$ in the low-frequency response region, following previous protocols for microgel-laden interfaces~\cite{Brugger2010} with one equilibration cycle and six measurement cycles per point, \partFig{fig:geometry}{d}. Strain was increased logarithmically at 20 points/decade from 0.001 to 1.0 strain amplitude and we report the strain-dependent elastic ($\elastic$) and loss ($\loss$) moduli from the primary Fourier components. The sinusoidal oscillation of the ring, $\theta(t) = \theta_0\sin(2\pi f t)$, is converted into an interfacial strain, $\gamma^s_0 = \theta_0 [(1-(R_{r,o}/R_{o})^2)^{-1}+((R_{r,i}/R_{i})^2-1)^{-1}]$ using the 2D-equivalent expressions for a Couette cylinder at the position of the ring~\cite{Macosko1994}. The strain can be approximated as $\gamma^s_0 \approx \theta R_{r,o}/(R_o-R_{r,o})$~\footnote{This can be recovered by factorising the denominator, $\gamma^s_0 = \theta_0 [R_o^2/[(R_o-R_{r,o})(R_{o}+R_{r,o})] + R_i^2/[(R_{r,i}-R_{i})(R_{r,i}+R_i)]$, and approximating to first order by taking $R_{i/o}+R_{r,i/o} \approx 2R_{r,i/o}$, $R_{r,i}\approx R_{r,o}$ and $R_o-R_{r,o} \approx R_{r,i}-R_i$.}, \ie\ the displacement of the ring divided by the distance from the ring to the outer pinned interface, analogous with the expression for the contactless geometry. As the ring is in direct contact with the interface, the measured torque can be converted straight to an interfacial stress, $\sigma^s = T/2\pi(R_{r,i}^2+R_{r,o}^2)$. All reported data is at high Bo, such that sub-phase drag correction is not applied, and low raw-phase angle, where geometry inertia does not dominate. Due to noise from the strain-control feedback loop, torque resolution is limited to \SI{0.07}{\micro\newton\metre}\cite{Renggli2020} and we highlight or truncate data below this threshold.

\subsection{Measuring stress in a contactless geometry}\label{sec:shear}
\begin{figure*}[tbp]
    \centering
    \includegraphics{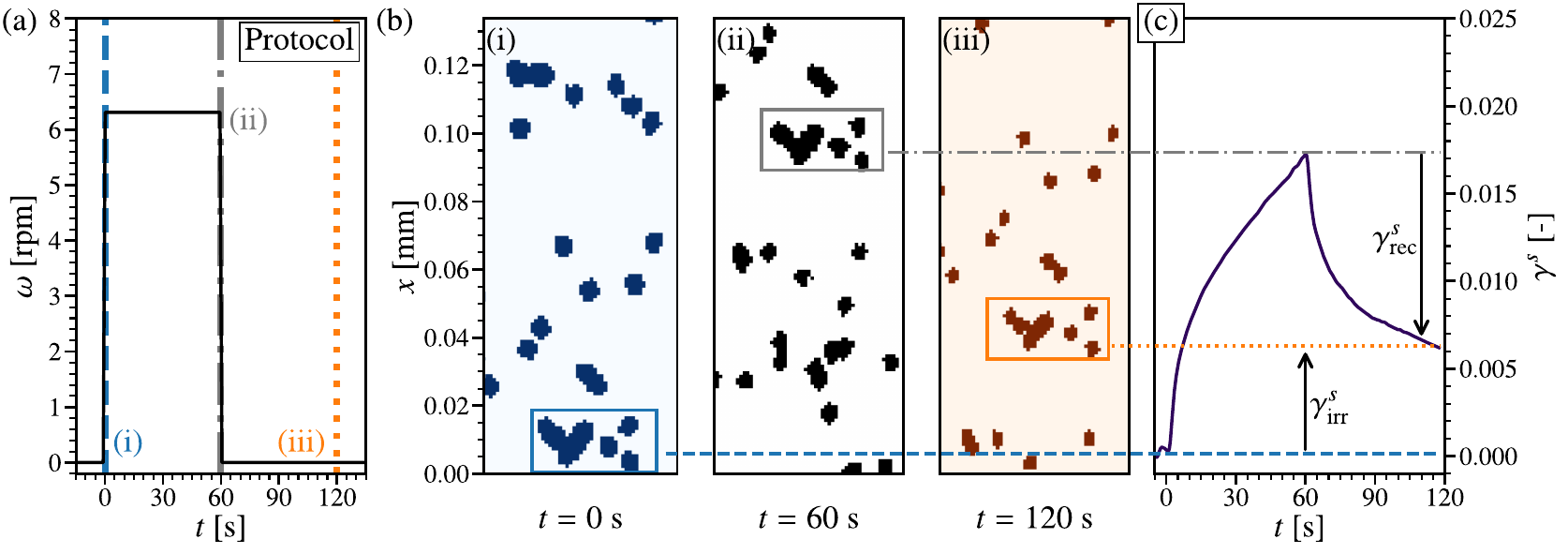}
    \caption{Contactless interfacial rheometry of \pnip--laden interfaces. (a)~Creep-recovery protocol, applied rotation rate, $\omega = 6.31$\,rpm, with time, $t$. Stress, $\sigma^s = \SI{3.2D-6}{\pascal\metre}$, for $\omega_i\approx 0$, Eq.~\eqref{eq:surfaceStress}. Highlighted $t$: (i)~dashed (blue), start of applied $\omega$ at $t=\SI{0}{\second}$; (ii)~dot-dashed (grey), end of applied $\omega$ at $t_{\omega} = \SI{60}{\second}$; and (iii)~dotted (orange), end of recorded recovery at $t=\SI{120}{\second}$. (b)~Measured interface response at highlighted $t$. Zoomed-in confocal microscopy at $t$ in (a). Outline, distinctive particle cluster showing interface motion along $x$, axis rotated to match (c). (c)~Extracted time-dependent strain, $\gamma^s(t)$. Horizontal lines trace strain from: (i)~dashed, (ii)~dot-dashed and (iii)~dotted. Arrows: recoverable (elastic) strain, $\gamma^s_{\rm rec} = \gamma^s(\SI{60}{\second}) - \gamma^s(\SI{120}{\second})$; and irrecoverable (plastic) strain, $\gamma^s_{\rm irr} = \gamma^s(\SI{120}{\second}) - \gamma^s(\SI{0}{\second})$.}
    \label{fig:protocol}
\end{figure*}

In contrast to the DWR geometry, the interface is not in direct contact with the geometry in the contactless method. Therefore, the stress cannot be directly measured by the rheometer. To make meaningful statements on the rheological properties of the interface, we must first find the interfacial stress applied at the interface. It is well known that for a parallel-plate setup, the stress is independent of the height through the sample upon reaching a steady state \cite{Macosko1994}. The timescale for momentum diffusion and to reach steady state is up to $\approx 0.3 h_o^2\rho_o/\eta_o = \mathcal{O}(\SI{1}{\second})$ for our setup~\cite{oza2021dynamics}, where $\rho_o$ is the oil density; this is far below the creep step duration. Considering the upper phase as a Newtonian fluid allows us to, therefore, find the stress on the interface from the upper fluid using the applied rotational speed of the geometry.

If the rheometer is rotated at a fixed angular velocity of $\omega$ then the shear rate of the upper fluid at a radius $r$ (Fig.~\ref{fig:torqueBalance}) of the parallel-plate geometry is given by
\begin{equation}
    \dot{\gamma} = \frac{\omega r}{h_o},
\end{equation}
where $h_o$ is the oil phase depth. This definition relies on the interface having zero speed. Therefore, this is a first approximation if the speed of the geometry is much larger than the speed of the interface. If this is not the case, then we can reduce $\omega$ by the angular speed of the interface at $r$, $\omega_i$, giving,
\begin{equation}
    \dot{\gamma} = \frac{(\omega-\omega_i) r}{h_o}.
\end{equation}

The stress induced is given by the product of $\dot\gamma$ with the upper phase bulk viscosity, $\eta_o$, \ie,
\begin{equation}
    \sigma = \frac{\eta_o (\omega - \omega_i)r}{h_o}.
\end{equation}
To convert this bulk stress into an interfacial stress we consider the torque applied from the bulk on an area element of a ring in the interface at $r$ of width ${\rm d}r$, Fig.~\ref{fig:torqueBalance}.

We can write the torque element, d$T$, as a product of the force element, $\sigma {\rm d}A$, and the radius, where d$A = 2\pi r{\rm d}r$ is the area of the infinitesimal ring,
\begin{equation}
    {\rm d}T = \sigma 2\pi r^2 {\rm d}r = 2\pi\frac{\eta_o(\omega - \omega_i)}{h_o} r^3 {\rm d}r.
    \label{eq:bulkTorque}
\end{equation}

This torque then gives rise to the interfacial stress, $\sigma^s$. With $T(r) = 2\pi r^2 \sigma^s(r)$ the interfacial torque as the product of $\sigma^s$, radius and perimeter, we can write the torque balance
\begin{equation}
    \begin{split}
        {\rm d}T = &T(r + {\rm d}r) - T(r) \\\approx &2\pi\left[(r+ {\rm d}r)^2\sigma^s(r+{\rm d}r) - r^2\sigma^s(r)\right]\\
        \approx& 2\pi r\left[r {\rm d}\sigma^s +2\sigma^s{\rm d}r  \right]
    \end{split}
    \label{eq:intTorque}
\end{equation}
where second order differential terms, \eg, $({\rm d}r)^2$, have been dropped. Equating Eqs~\eqref{eq:bulkTorque} and \eqref{eq:intTorque} and rearranging,
\begin{equation}
    \frac{{\rm d}\sigma^s}{{\rm d}r} + \frac{2\sigma^s}{r} = \frac{\eta_o(\omega-\omega_i)}{h_o}r.
    \label{eq:stressDiff}
\end{equation}

\begin{figure*}[tbp]
    \centering
    \includegraphics{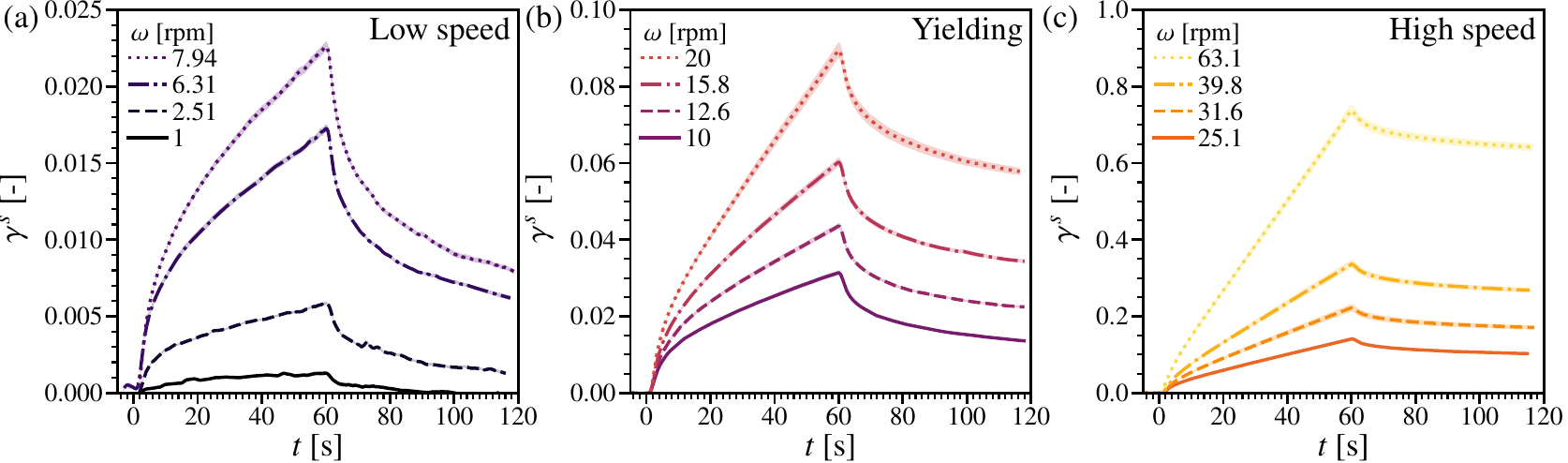}
    \caption{Strain response, $\gamma^s(t)$, of PNIPAM-SiO$_2$--laden interface for contactless rheology protocol, \partFig{fig:protocol}{a}, showing transition from elastic dominated response to plastic flow with increasing rotation rate, $\omega$. (a)~Low $\omega = \SIrange{1}{8}{rpm}$ (dark to light), see legend, with dominant elastic response, $\gamma^s_{\rm rec}$, after $t_{\omega} = \SI{60}{\second}$. Shading, error from standard deviation in image correlation analysis bands. (b)~Moderate $\omega$ around yielding with rising plastic response. (c)~Strain response at high $\omega$ dominated by irreversible plastic flow, $\dot\gamma^s$ (see text for details).}
    \label{fig:microgelStrain}
\end{figure*}
Equation~\eqref{eq:stressDiff} can be readily solved to give
\begin{equation}
    \sigma^s = \frac{\eta_o(\omega-\omega_i)}{4h_o}r^2
    \label{eq:surfaceStress}
\end{equation}
if $\frac{{\rm d}\omega_i}{{\rm d} r}\simeq 0$, so there is no interfacial shear banding, for example; this aligns with our confocal-microscopy observations, \eg, $\omega_i$ is constant in time and across the field of view (within error). It is evident this expression yields the correct dimensions for interfacial stress as Pa\,m, as well as physically reasonable dependencies on viscosity, applied rotation, oil phase height, and radius. In order to vary the interfacial stress, we vary $\omega$ while observing at a fixed $r$, a practically simpler approach.

Finally, we need to ensure that we are measuring surface rather than bulk sub-phase properties. Typically, this can be checked via Bo [Eq.~\eqref{eq:boussinesq}], but this assumes a probe in contact with the interface. For our contactless setup, as we measure interfacial strain via the motion of the interface itself, we focus instead on the question: is the force on the \emph{interface} dominated by surface or bulk sub-phase viscosity? This effect is included in sub-phase drag corrections~\cite{Verwijlen2011}, but is typically neglected in Bo as drag on the probe dominates. For our contactless geometry, the stress from the sub-phase can be crudely approximated as parallel-plate flow with $\omega_i$, giving an equivalent to Bo,
\begin{equation}\label{eqn:boussinesq_particle}
    \mathrm{Bo}^* = \frac{\sigma^s}{\sigma^{s}_{\rm drag}} = \frac{\eta_o h_w}{\eta_w h_o}\frac{\omega-\omega_i}{\omega_i} \, \mathrm{for}~ \sigma^{s}_{\rm drag} \approx \frac{\eta_w\omega_i}{4h_w}r^2,
\end{equation}
where $\eta_w$ and $h_w$ are the water sub-phase viscosity and depth, \partFig{fig:geometry}{a}. The first term in Bo$^*$ is $\mathcal{O}(1)$ for our bulk phases and dimensions, and likely most common uses, but the second term can be arbitrarily large as $\omega_i \rightarrow 0$. This makes the technique suitable for measuring weak interface yielding, a fact that arises from decoupling stress application and interface motion by not having a direct probe. For interfaces with significant $\omega_i$, Bo$^*$ can drop and this is discussed where relevant. 

\section{Results and Discussion}

\subsection{\texorpdfstring{Elastic PNIPAM-SiO$_2$--laden interface}{Elastic PNIPAM-SiO2--laden interface}}

To establish the validity of our novel contactless interfacial rheometric technique we begin by measuring a highly elastic PNIPAM-SiO$_2$--laden interface, which can also be studied by conventional DWR interfacial rheometry.

\subsubsection{Contactless rheometry}

Using the contactless setup, we perform a creep-recovery test with increasing rotation rates, $\omega$, from 0.1 to 400~revolutions per minute (rpm) with logarithmic spacing at 5~pts/decade and higher resolution (20~pts/decade) where behaviour is observed to be changing from the confocal microscopy recordings. At each step, while recording, $\omega$ is applied for \SI{60}{\second} before zero rotation rate is set for a further \SI{60}{\second}, \partFig{fig:protocol}{a}. Imaging tracer particles embedded in the interface throughout these steps gives the resulting deformation of the interface. This is illustrated by following a distinctive cluster of particles in magnified images, \partFig{fig:protocol}{b}. From the start of applied rotation (i) to when rotation is stopped (ii), the interface first moves along the flow direction, $x$. After cessation and relaxation the interface recoils backwards, along the previous flow direction, until the recovery step ends, \partFig{fig:protocol}{b)(ii) to (iii}. Over time, this particle motion can be seen as tracing out the strain response of the interface, $\gamma^s(t) = x(t)/r_{\rm out}$, lines from (b)(i)--(iii) to (c).   

At low applied $\omega$, \eg, $6.31$\,rpm (Fig.~\ref{fig:protocol}),
there is an initial jump in $\gamma^s$ as $\omega$ is applied before a further slow increase over $t_{\omega}$ [the creep step length, \partFig{fig:geometry}{d}], see \partFig{fig:protocol}{c}. At the cessation of applied flow there is a rapid recoil, followed by a slower further relaxation, approaching a constant value. From the strain profile, $\gamma^s(t)$, we extract two quantities: the recoverable strain as the decrease from the peak to the final strain [dot-dashed to dotted lines, (ii) to (iii)], and the irrecoverable strain, $\gamma^s_{\rm irr}$, as the increase from the initial strain, at the start of shear, to the final strain [dashed to dotted lines, (i) to (iii)]. A strong initial deformation and near complete elastic recovery can be seen over a range of $\omega \lesssim 8$\,rpm, \partFig{fig:microgelStrain}{a}, with both steps growing with $\omega$ (dark to light).

As $\omega$ increases further, up to $20$\,rpm [\partFig{fig:microgelStrain}{b} (dark to light)], $\gamma^s_{\rm rec}$ increases proportionally; $\gamma^s_{\rm irr}$ also begins to slowly increase as the interface does not fully recover. At higher $\omega$ still, $\gtrsim 25$\,rpm [\partFig{fig:microgelStrain}{c}]  there is a clear change in behaviour, as $\gamma^s_{\rm irr}$ rapidly increases while the elastic recovery remains unchanged. During applied rotation, the strain is linear in time, giving a well-defined interfacial shear rate, $\dot\gamma^s = \gamma^s_{\rm irr}/t_{\omega}$. This behaviour is indicative of yielding in a creep-recovery test~\cite{Nguyen1992}.

\begin{figure}[tb]
    \centering
    \includegraphics{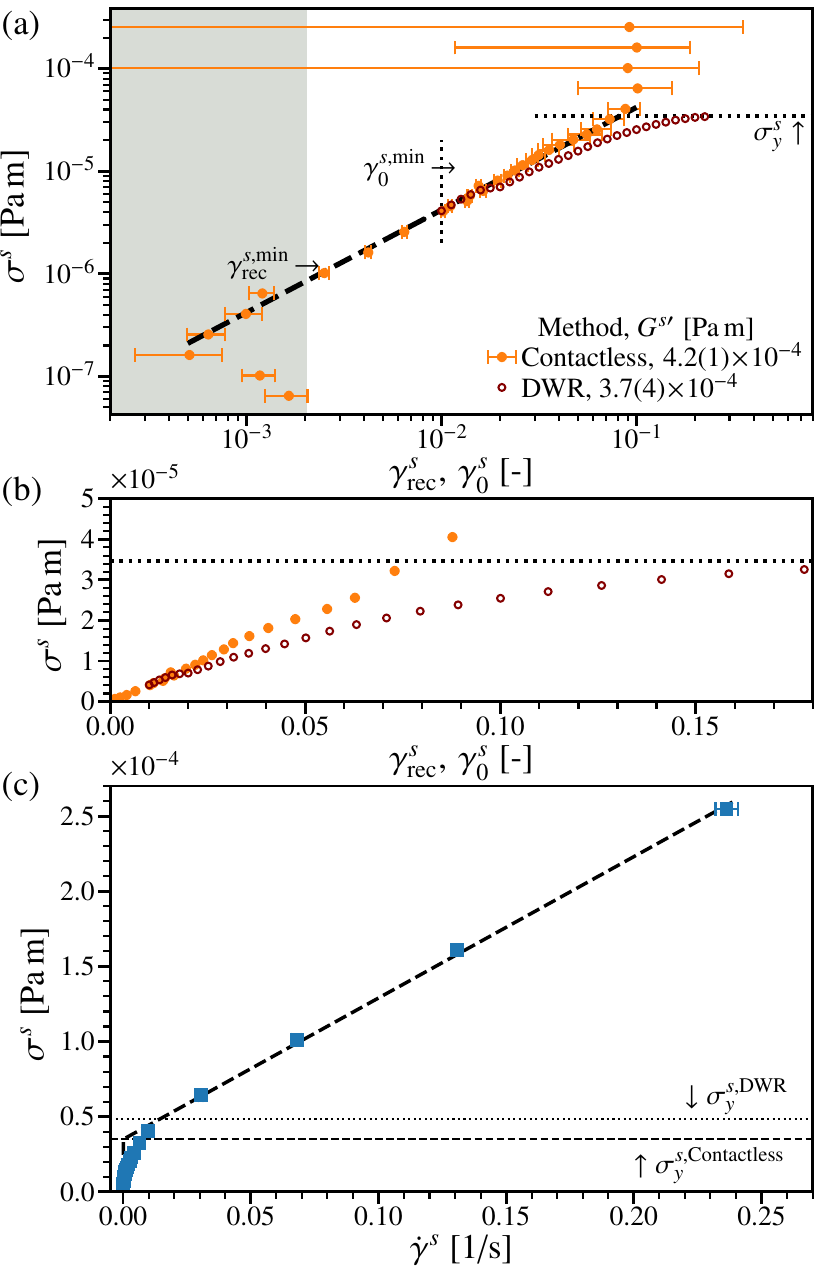}
    \caption{PNIPAM-SiO$_2$--laden interface rheology. (a)~Elastic response, stress ($\sigma^s$) vs strain. Points: solid, contactless [$\sigma^s$ from Eq.~\eqref{eq:surfaceStress} and strain, $\gamma^s_{\rm rec}$, from relaxation, Fig.~\ref{fig:microgelStrain}]; open, DWR elastic stress, [$\sigma^s = \gamma^s_0\elastic$ for strain amplitude, $\gamma^s_0$]. Minimum limits: shading, contactless strain, $\gamma_{\rm rec}^{s,\min}\! \approx\! 0.002$; vertical dotted line, DWR torque at $\gamma_0^{s,\min}$. Fit lines: dashed, contactless elastic response, $\elastic \!=\! \sigma^s/\gamma^s_{\rm rec} \!= \!\SI{4.2(1)D-4}{\pascal\metre}$; horizontal dotted, yield stress from (c). Fit of linear elastic response for DWR in Fig.~\ref{fig:DWRrheo}. (b)~Linear plot of elastic response before yielding, symbols as in (a). (c)~Contactless viscous response. Points, stress vs shear rate, $\dot\gamma^s = \gamma^s_{\rm irr}/t_{\omega}$ from irrecoverable strain, \partFig{fig:microgelStrain}{c}. Lines: bold dashed, Bingham plastic fit, Eq.~\ref{eq:bingham}; fine dashed, yield stress, $\sigma_y^{s,\rm Contactless}\! =\! \SI{3.5D-5}{\pascal\metre}$; and dotted, DWR yield stress, $\sigma_y^{s,\rm DWR}\! =\! \SI{4.8D-5}{\pascal\metre}$.}
    \label{fig:microgelRheology}  
\end{figure}

Using $\gamma^s_{\rm rec}$ and $\dot\gamma^s$ alongside the calculated interfacial stress ($\sigma^s$), Eq.~\eqref{eq:surfaceStress}, the rheological response can be quantified. The elastic response, $\sigma^s(\gamma^s_{\rm rec})$, shows three regimes, \partFig{fig:microgelRheology}{a}~[solid circles]. At the lowest stresses, $\sigma^s<\SI{D-6}{\pascal\metre}$, no clear trend is observed. This (shaded) region lies below a minimum strain, $\gamma_{\rm rec}^{s,\min} \approx 2\times 10^{-3}$, set by noise, \eg, vibration or drift. With increasing stress, $\gamma^s_{\rm rec}$ increases linearly until $\sigma^s = \SI{5D-5}{\pascal\metre}$. Above this the elastic recovery appears constant, but noisy (\ie\ spatially heterogeneous across the field of view). Within the linear region an elastic constant, $\elastic = \sigma^s/\gamma^s_{\rm rec} = \SI{4.2(1)D-4}{\pascal\metre}$ can be fitted (dashed line). The interface is well described as a linear elastic solid below $\sigma^s = \SI{5D-5}{\pascal\metre}$, this can be emphasised by plotting on linear axes, \partFig{fig:microgelRheology}{b}.

However, this is only one side of the measured response. The stress-dependent plastic flow, $\sigma^s(\dot\gamma^s)$, further illuminates the change around $\sigma^s \approx \SI{5D-5}{\pascal\metre}$, \partFig{fig:microgelRheology}{c}~(squares). Below this threshold $\dot\gamma^s$ is limited, but on further increase the interface begins to flow $\dot\gamma^s \sim \mathcal{O}(\SI{0.1}{\per\second})$. The transition point, or yield stress ($\sigma^s_y$) can be quantified by fitting a simple piecewise function to these data. We model the interface as a Bingham fluid, one of the simplest models capturing suitable non-Newtonian behaviour, described by:
\begin{equation} \label{eq:bingham}
  \begin{array}{lr}
    \dot{\gamma}^s = 0 & : \sigma^s < \sigma^s_y\\
    \sigma^s = \sigma^s_y + \eta^s\dot\gamma^s & : \sigma^s \ge \sigma^s_y.
  \end{array}
\end{equation}
Below $\sigma^s_y$ there is no flow, above this the excess stress leads to a shear rate set by the interfacial viscosity, $\eta^s$. From this we can obtain both a yield stress and an interfacial viscosity from the contactless technique, $\sigma^{s,\rm Contactless}_y = \SI{3.5(1)D-5}{\pascal\metre}$, and $\eta^s = \SI{9.4(1)D-4}{\pascal\metre\second}$. While the Bingham model is appropriate to capture a clear yield stress transition, in general more complex interfacial yielding behaviour is often observed~\cite{erni2012nonlinear,truzzolillo2016tuning}.

\subsubsection{Comparison to DWR rheometry\label{sec:microgelDWR}}

\begin{figure}[tb]
    \centering
    \includegraphics{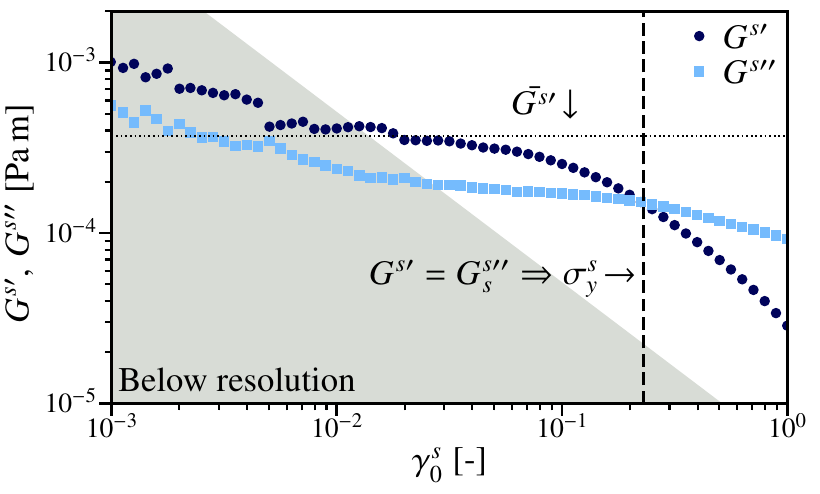}
    \caption{DWR rheology of a PNIPAM-SiO$_2$--laden interface. Elastic ($\elastic$, dark circles) and loss ($\loss$, light squares) moduli vs strain amplitude, $\gamma^s_0$. Shading, below instrument resolution. Lines: dotted, $\bar{\elastic} = \SI{3.7(4)D-4}{\pascal\metre}$, mean of $\elastic(\gamma^s_0 \leq 0.05)$; dashed, yielding at $\elastic=\loss$, $\sigma^{s,\rm DWR}_y = \gamma^s_0 \sqrt{(\elastic)^2+(\loss)^2} = \SI{4.8D-5}{\pascal\metre}$}
    \label{fig:DWRrheo}
\end{figure}

As a strong and highly elastic interface we can directly compare the results of the contactless geometry to DWR interfacial rheology for the same PNIPAM-SiO$_2$ at equal surface pressures. As the torque resolution for oscillatory tests is finer than steady shear for rotational rheometers, an oscillatory strain amplitude sweep is performed. This measures the strain-dependent elastic and loss moduli, Fig.~\ref{fig:DWRrheo} (symbols). At low $\gamma^s_0<0.1$, the elastic modulus is higher than the loss modulus, with $\elastic$ only weakly decreasing once above the torque resolution (shaded region), indicative of a solid elastic material as the stress in phase with the strain. With increasing $\gamma^s_0$, $\elastic$ begins to drop while $\loss$ remains near constant. At $\gamma^s_0=0.23$ the moduli become equal, $\elastic = \loss$; above this point $\elastic$ continues to sharply drop, while $\loss$ weakly decreases. In this region the stress is in phase with the shear rate, \ie\ liquid-like. Therefore, with increasing strain amplitude the interface yields from a solid-like state to a liquid-like state where $\elastic = \loss$. This stress, $\sigma^{s,\rm DWR}_y = \SI{4.8D-5}{\pascal\metre}$, is an operative yield stress at a finite frequency and shear rate, in contrast to the `static' measurement of creep recovery~\cite{dinkgreve2016different}.

The comparison of $\sigma_y^{s,\rm DWR}$ [\partFig{fig:microgelRheology}{c} (dotted line)] with the contactless rheology value, $\sigma_y^{s,\rm Contactless}$ (fine dashed line), finds similar values, with only a 30\% difference, 4.8 vs \SI{3.5D-5}{\pascal\metre}. This is within the expected variation for different measurement protocols of a non-linear property~\footnote{Oscillatory yielding is ambiguous, as it can be gradual, with multiple definitions; $\elastic=\loss$ is an upper (over)estimate~\cite{dinkgreve2016different}. \textit{E.g.}, a tangent analysis appears to give better agreement with the contactless method, but requires extrapolation from data below instrument resolution.}, \eg, in colloidal gels or glasses~\cite{Pham2008}. The yield strains, $\gamma^s_0 =0.23$ and $\gamma^s_{\rm rec}=0.1$, are also comparable, \partFig{fig:microgelRheology}{a}, but greater than previously observed for microgel-laden~\cite{Brugger2010} or amorphous jammed interfaces~\cite{Galloway2020}. Above yielding we are not able to compare DWR and contactless measurements, as the applied shear rates for DWR ($\dot\gamma^s = 2\pi f \gamma^s_0 \gtrsim \SI{0.25}{\per\second}$) are larger than those in the contactless geometry. To yield at comparable $\dot\gamma^s \approx \SI{0.01}{\per\second}$ would require $f\approx\SI{0.01}{\hertz}$, resulting in infeasibly long experiments.

\begin{figure*}
    \centering
    \includegraphics[height=0.28\textwidth,trim={0cm -3cm 0.0cm 0.0cm},clip]{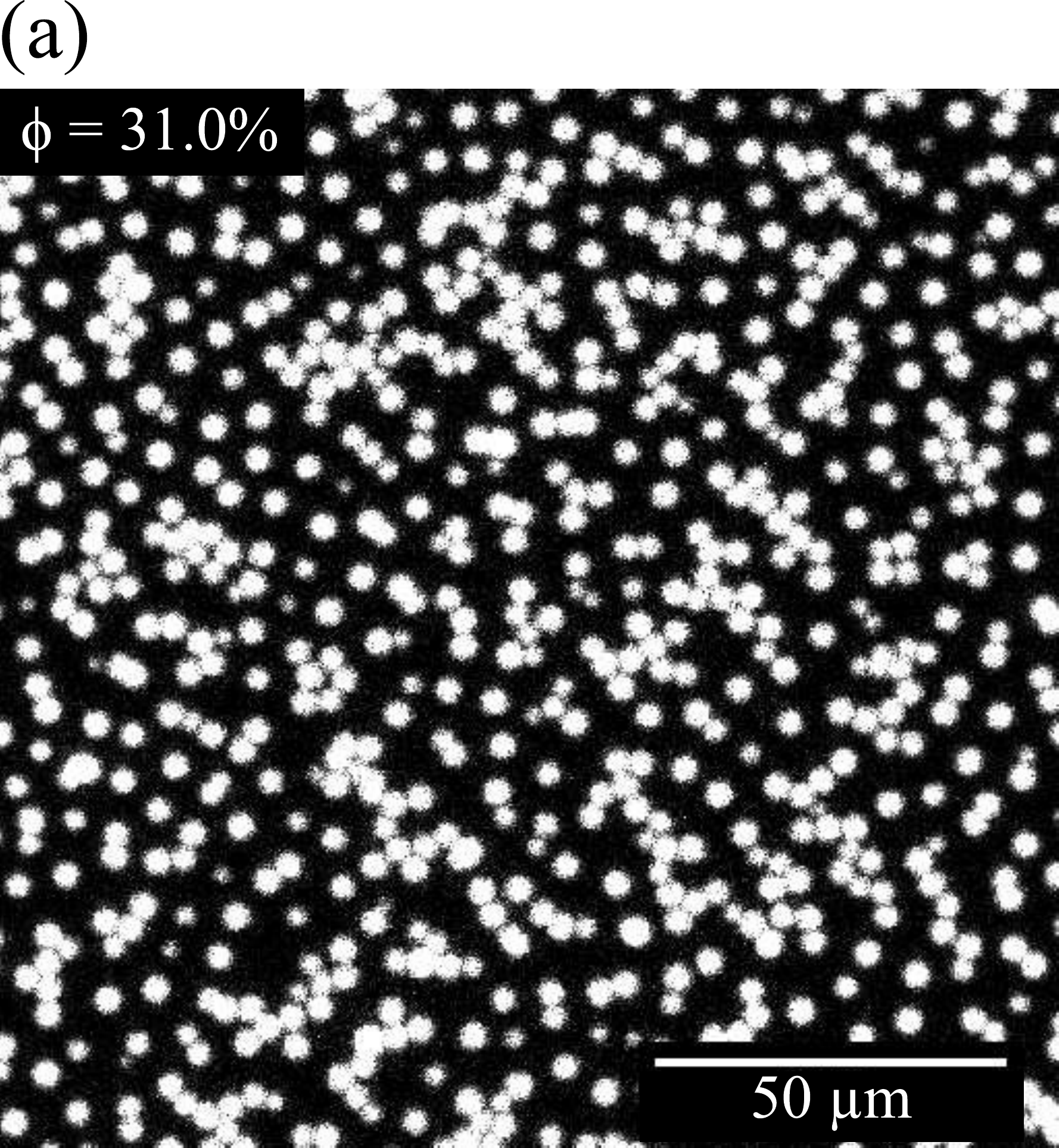}
    \includegraphics[height = 0.28\textwidth]{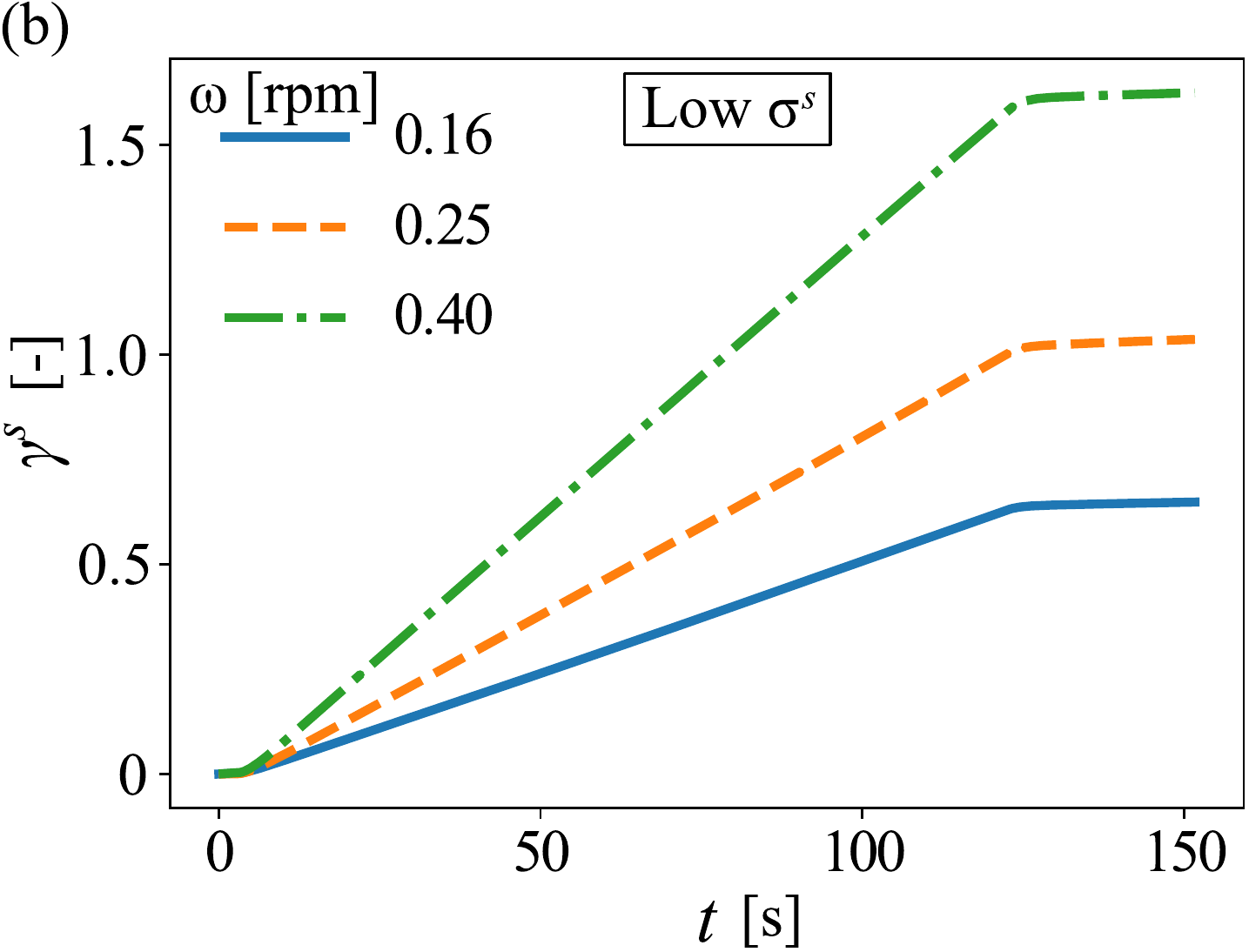}
    \includegraphics[height = 0.28\textwidth]{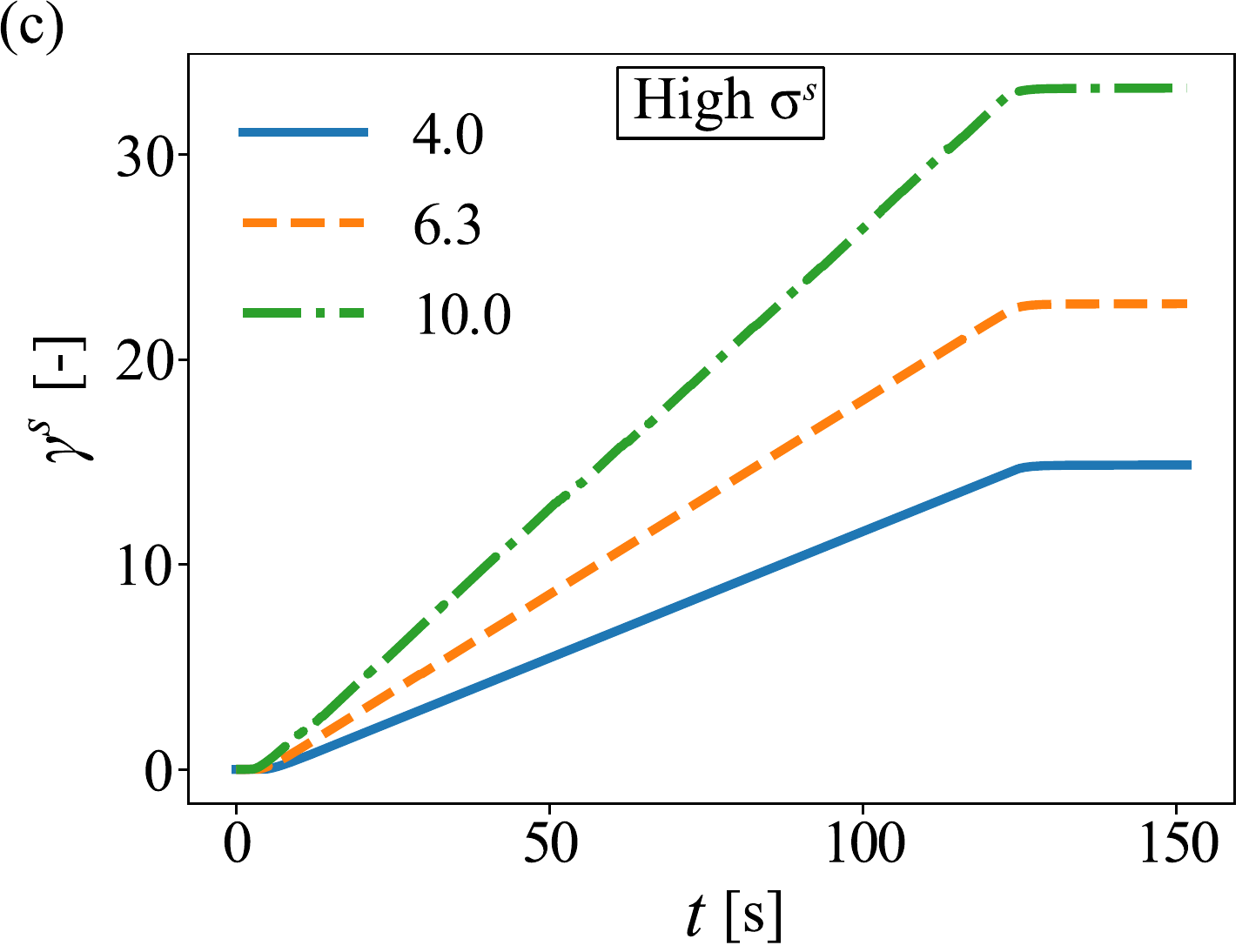}
    \caption{PMMA-laden interface at low surface coverage. (a)~Fluorescent confocal micrographs of PMMA particles (white) at an oil-water interface with $\phi= 31.0\%$. (b)~Corresponding strain vs time, $\gamma^s(t)$, at low imposed rheometer rotation rates, $\omega$, see inset legend. $r_{\rm out}=\SI{6.6}{\milli\metre}$. $\omega$ applied from $t\gtrsim \SI{5}{\second}$ for \SI{120}{\second}, followed by \SI{30}{\second} further recording. (c)~Corresponding high imposed $\omega$ response.}
    \label{fig:lowPhi}
\end{figure*}

In contrast to the yield stress, the linear elastic behaviour is well-defined. Comparing the DWR elastic modulus below yielding, \partFig{fig:microgelRheology}{a} (open circles), and the contactless elastic modulus (filled circles) we see excellent agreement. Fitting over the linear response region ($\gamma^s_0 \leq 0.05$) the average elastic modulus from DWR is only 12\% lower than the contactless measurement, $\bar{\elastic} = 3.7(4)$ vs \SI{4.2(1)D-5}{\pascal\metre}, demonstrating that they measure equal quantities within error. As yielding is approached, the DWR response appears to soften while the contactless measurement remains linear, \partFig{fig:microgelRheology}{b}. However, the oscillatory strain amplitude will contain both the recoverable elastic strain and plastic flow~\cite{Kamani2021}, in contrast to a creep-recovery measurement that separates the terms. So, approaching yielding, where plastic flow gradually begins [\partFig{fig:microgelRheology}{c}] they are no longer directly comparable. The comparable $\sigma^s_y$ values and equal elastic moduli demonstrate that the contactless geometry accurately measures well-defined interfacial rheological properties.

\subsection{PMMA particle laden interface\label{sec:PMMAresults}}

\subsubsection{Low surface coverage}

We now turn to an interface laden with solid colloidal hydrophobic PMMA particles, which are typically challenging to measure using conventional interfacial rheometric techniques~\cite{VanHooghten2017}. First, we investigate interfaces with low surface fraction, e.g.~$\phi = 31$\% [\partFig{fig:lowPhi}{a}]. Interestingly, even though the particles are hydrophobic sterically stabilised nearly hard-sphere particles, they exhibit long-range repulsion when confined at an oil-water interface and assemble predominantly in a non-close packed arrangement \cite{Muntz2020}. 
Some aggregation is present due to attractive capillary and Van der Waals forces, \partFig{fig:lowPhi}{a}. While capillary forces should be negligible, due to a vanishingly small Bond number and the use of spherical particles~\cite{Binks2006}, there will be a certain particle roughness from the variable length of the steric stabiliser ``hairs''. This may cause contact line undulations that lead to short-range capillary attraction~\cite{Stamou, Hooghten}.  

For this particle system, strain vs time plots show smooth flow with a constant shear rate over the duration of the creep step, \partFig{fig:lowPhi}{b}.  As expected, at larger imposed rotation rates from the rheometer a larger interfacial shear rate is measured in response, \partFig{fig:lowPhi}{c}. Note that, when shear starts the interface appears to immediately (within temporal resolution of the analysis method) begin flowing at a constant shear rate. Similarly, when the shear ends the interface immediately stops flowing. This implies that the interface response in this regime is purely viscous, with no measurable elasticity. 

\begin{figure}
    \centering
    \includegraphics[width = 0.99\columnwidth]{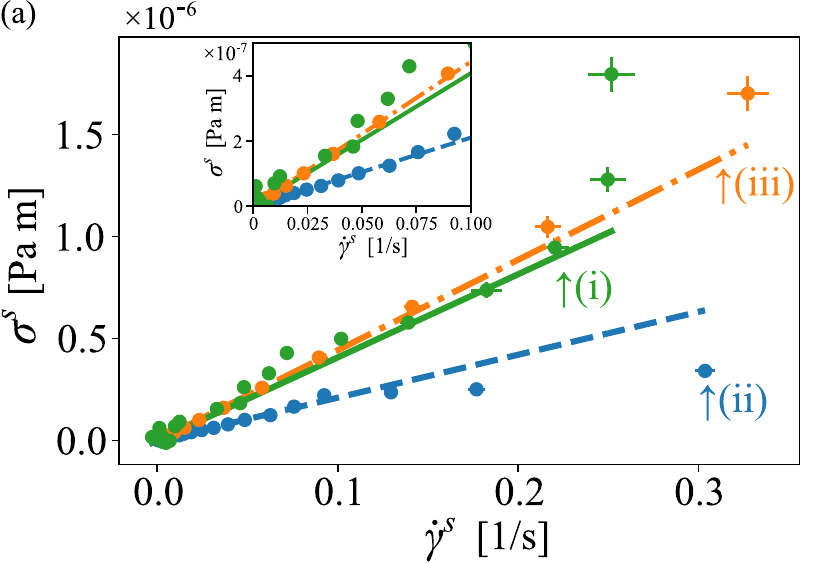}\\        \includegraphics[width=0.32\columnwidth]{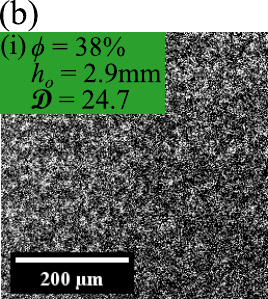}
    \includegraphics[width=0.32\columnwidth]{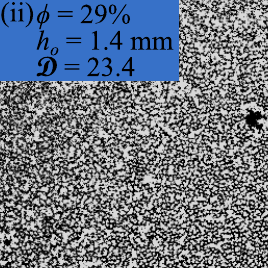}
    \includegraphics[width=0.32\columnwidth]{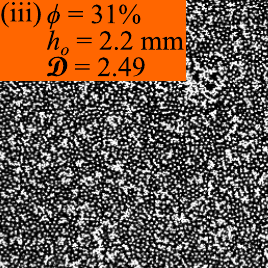}
    \caption{Newtonian rheological behaviour of low surface coverage interfaces with varying parameters. (a)~Stress--shear-rate behaviour of three interfaces with varying oil heights ($h_o$), surface coverages ($\phi$) and aggregation states ($\mathcal{D}$). Points, measured data. Lines, fits to constant viscosity: (i)~\SI{4.1(5)D-6}{\pascal\metre\second}, solid (green); (ii)~\SI{2.10(10)D-6}{\pascal\metre\second}, dashed (blue); (iii)~\SI{4.43(9)D-6}{\pascal\metre\second}, dot-dashed (orange), also in Fig.~\ref{fig:lowPhi}. (b)~Interface images: (i)~$\phi=38\%$, $h_o=\SI{2.9}{\milli\metre}$ and $\mathcal{D} =24.7$; (ii)~$\phi=29\%$, $h_o=\SI{1.4}{\milli\metre}$ and $\mathcal{D} = 23.4$; and, (iii)~$\phi=31\%$, $h_o=\SI{2.2}{\milli\metre}$ and $\mathcal{D}=2.49$. NB: surface fractions are determined from a rotational average, so single images are not wholly representative.}
    \label{fig:oilThickness}
\end{figure}

To quantify the rheology of these particle-laden interfaces, we plot stress vs shear rate, $\sigma^s(\dot\gamma^s)$, with $\dot\gamma^s$ defined from the slope of $\gamma^s(t)$, Sec.~\ref{sec:mat_and_methods}, due to the immediate response. As expected, for relatively low $\phi$ we observe a Newtonian response, \partFig{fig:oilThickness}{a}~[dot-dashed (orange) line]. We can therefore assign a constant $\eta^s = \SI{4.43(9)D-6}{\pascal\metre\second}$ for the interfacial viscosity of this interface. This $\eta^s$ is comparable to that measured using a magnetic rod interfacial rheometer on a similar system~\cite{Reynaert2007}. 

\paragraph{Effect of particle assembly.\label{sec:low_surfAgg}}

Next, we have performed repeats of these experiments while varying the oil phase thickness to test the robustness of our technique, Fig.~\ref{fig:oilThickness}. Confocal images prior to shearing reveal that the partial presence of aggregates, with particles in direct contact, varies between samples, \partFig{fig:oilThickness}{b}. The aggregation state is characterized by the dispersity $\mathcal{D}$, Eq.~\eqref{eq:pd}, where a low $\mathcal{D}$ corresponds to a homogeneous particle distribution and a high $\mathcal{D}$ to an aggregated assembly. We will discuss the particle assembly's influence on the rheology below.

First, we observe a Newtonian response for all three oil thicknesses, \partFig{fig:oilThickness}{a}. This suggests that, as expected, the height of the oil phase does not seem to have a substantial effect, \ie\ the rheology of the interface is not expected to depend on the depth of the bulk phases. When comparing samples with different $\phi$ and aggregation states, we observe the following trends. First, an increase in $\phi$ leads to an increase in interfacial viscosity. For example, when comparing interfaces with similar aggregation states 
($\mathcal{D}=24.7$ and $\mathcal{D}=23.4$), 
but different $\phi$, the interfacial viscosity at 38\%, (green) solid line, is higher than at 29\%, (blue) dashed, 4.1 vs \SI{2.1D-6}{\pascal\metre\second}. This trend is to be expected, as we later demonstrate elastic responses for high surface fractions of 
56.7\%, Sec.~\ref{sec:high_surf}. Second, the aggregation state seemingly affects the interfacial viscosity. When comparing samples with similar $\phi$ (29\% and 31\%), but different aggregation states ($\mathcal{D}=23.4$ and $\mathcal{D}=2.49$ respectively), we measure a significant difference in $\eta^s$ (2.1 vs \SI{4.4D-6}{\pascal\metre\second}), cf.\ (blue) dashed and (orange) dot-dashed lines. This suggests that the aggregation of PMMA particles at liquid interfaces leads to lower interfacial viscosities compared to more homogeneously distributed particles.

Our observation that surface coverage and aggregation state make a significant difference to measured interfacial viscosity demonstrates the utility of the inherent combination of rheometric measurement with simultaneous imaging for the contactless geometry. It also aligns with previous reports; for example, \citet{Reynaert2007}~have shown that the complex surface viscosity magnitude increases with the surface fraction of weakly aggregated polystyrene particles at a water-oil interface. They also show that, for surface coverages below 80\%, as studied here, the interfacial viscosity of aggregated particles at a water-oil interface is lower than that for stable particles. Note that aggregation \emph{can} lead to higher interfacial viscosities at high(er) surface coverage, see also, \eg, Ref.~\onlinecite{Barman2016}.

Notably, the results of varying the thickness of the oil phase also imply that edge effects, \ie\ deviations from our assumed parallel-plate flow field, Sec. \ref{sec:shear}, do not play a substantial role in our setup. This is consistent with the agreement in the measured linear elastic modulus for \pnip--laden interfaces between DWR and contactless methods, Sec.~\ref{sec:microgelDWR}. A substantial part of the top surface in our set-up is an oil-plate interface, but the outer edge of it is an oil-air interface \ie\ the sample cup is \SI{21}{\milli\metre} radius at the top, whereas the parallel-plate geometry attached to the oil-air interface is \SI{12.5}{\milli\metre} radius, leaving a radial gap of \SI{9}{\milli\metre} between the parallel-plate geometry and the inner wall of the PTFE cup. To mitigate any edge effects, we measure interfacial strains at a distance of about 4 mm $\ll$ 12.5 mm from the rotational axis. The results in Fig.~\ref{fig:oilThickness}, \ie\ that the interfacial viscosities do not differ strongly with oil-phase thickness, imply that edge effects do not substantially affect our results. However, we should note that even in the contactless method, at low $\phi$, Fig.~\ref{fig:lowPhi}, such low interfacial viscosities mean that measurements are at a moderate Bo, around $\mathcal{O}(10)$. This suggests that for precise and absolute determination of $\eta^s$ in this regime a sub-phase drag correction may still be necessary.

\begin{figure*}
    \centering
    \includegraphics{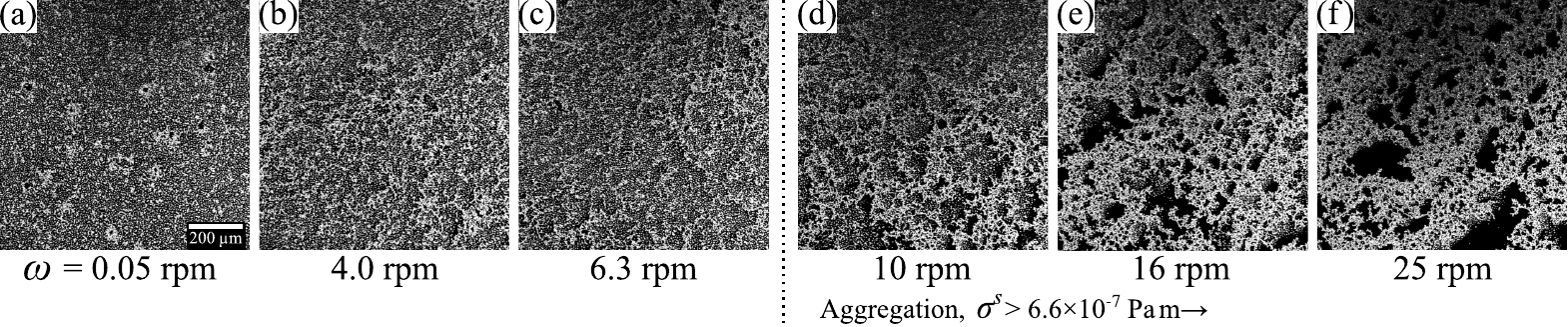}
    \caption{Structural evolution of PMMA particles at an oil-water interface with 31\% surface coverage via fluorescent confocal micrographs with increasing applied rotation rate, $\omega$, and, hence, interfacial stress, $\sigma^s$. Interface corresponding to (orange) dot-dashed line in \partFig{fig:oilThickness}{a}. (a)~After low stress, $\omega = 0.05$\,rpm. Scale bar \SI{200}{\micro\metre}. (b)~$\omega = 4.0$\,rpm. (c)~$\omega = 6.3$\,rpm. (d)~Aggregation threshold, $\omega = 10$\,rpm or $\sigma^s = \SI{6.6D-7}{\pascal\metre}$. (d)~Continued aggregation, $\omega = 16$\,rpm. (e)~Highest $\omega=25$\,rpm.}
    \label{fig:prepost}
\end{figure*}

\paragraph{Irreversible effect of shear.} 
Next, we take advantage of the simultaneous confocal imaging to observe the evolution in particle assembly upon shearing. Before shearing, the particles are mostly in a non-close packed arrangement with partial aggregation ($\mathcal{D}=2.49$), which we attribute to attractive capillary and Van der Waals forces, \partFig{fig:prepost}{a}. Upon mild shearing ($\omega \leq 6.3$\,rpm), the arrangement is preserved and only minor changes in aggregation state are observed, \partFig{fig:prepost}{b) and (c}. We see from images taken after high shear is applied, $\omega \gtrsim 10$\,rpm corresponding to $\sigma^s =\SI{6.6D-7}{\pascal\metre}$ [\partFig{fig:prepost}{d) and (e}], that the interface changes to an inhomogeneous structure with most particles forming one large aggregate percolating across the region imaged combined with an increase in dispersity to $\mathcal{D} = 13.5$ in the final state. Importantly, the aggregation appears irreversible and persists even when higher shear rates are applied, \partFig{fig:prepost}{f}.

\begin{figure*}
    \centering
        \includegraphics[height=0.28\textwidth,trim={0cm -3cm 0.0cm 0.0cm},clip]{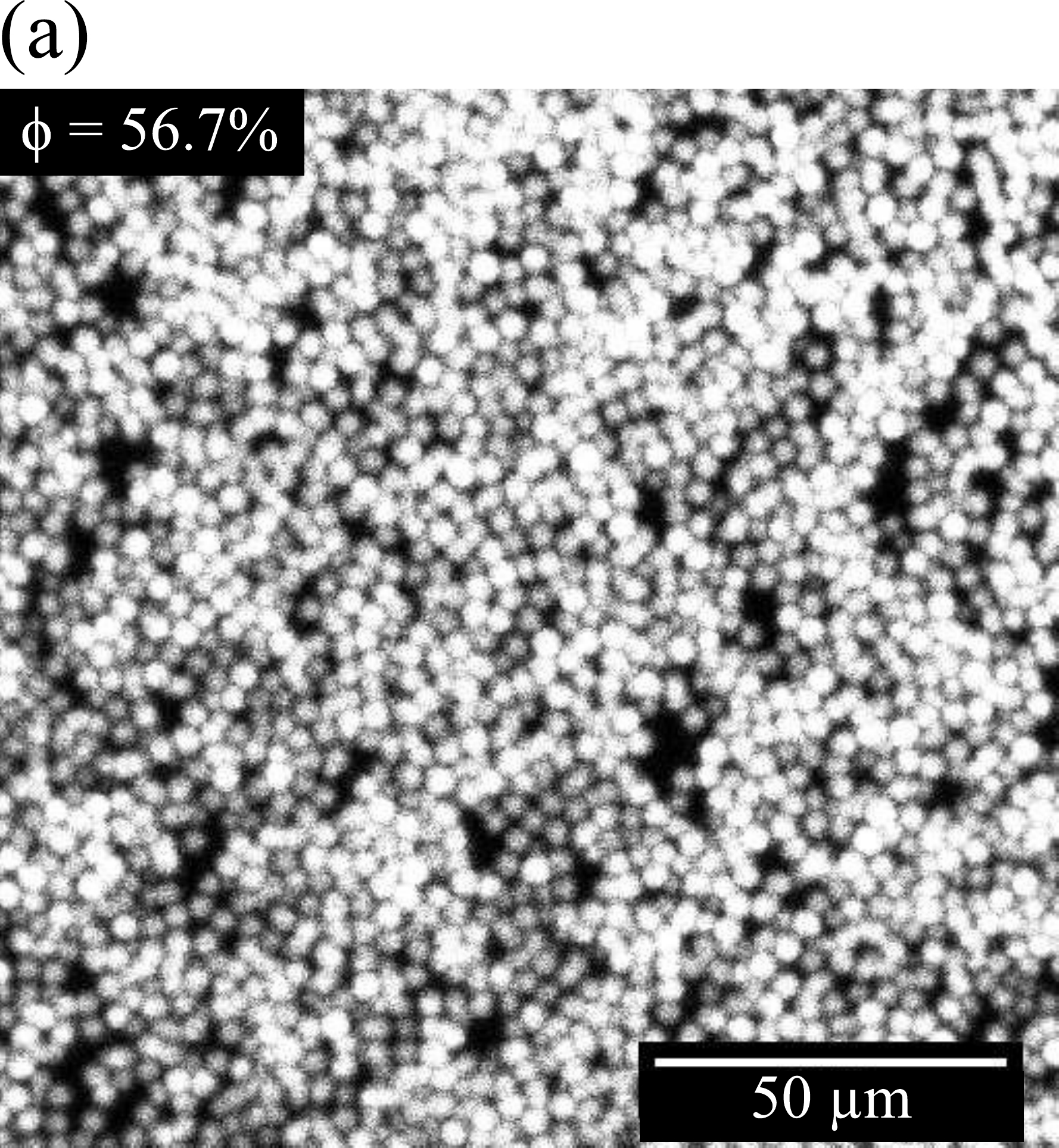}
        \includegraphics[height = 0.28\textwidth,trim={0.0cm 0cm 0.0cm 0.0cm},clip]{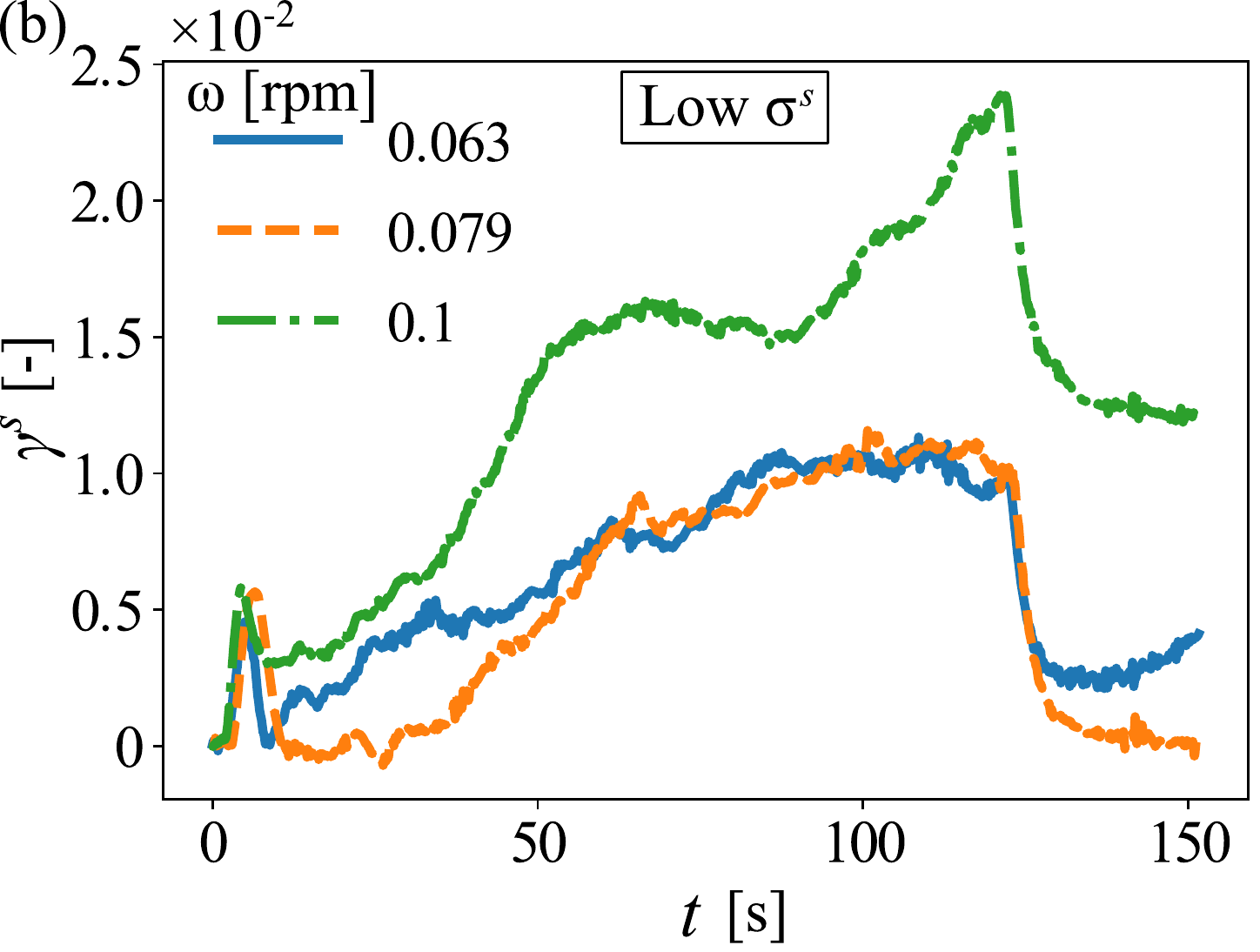}
        \includegraphics[height = 0.28\textwidth,trim={0.0cm 0cm 0.0cm 0.0cm},clip]{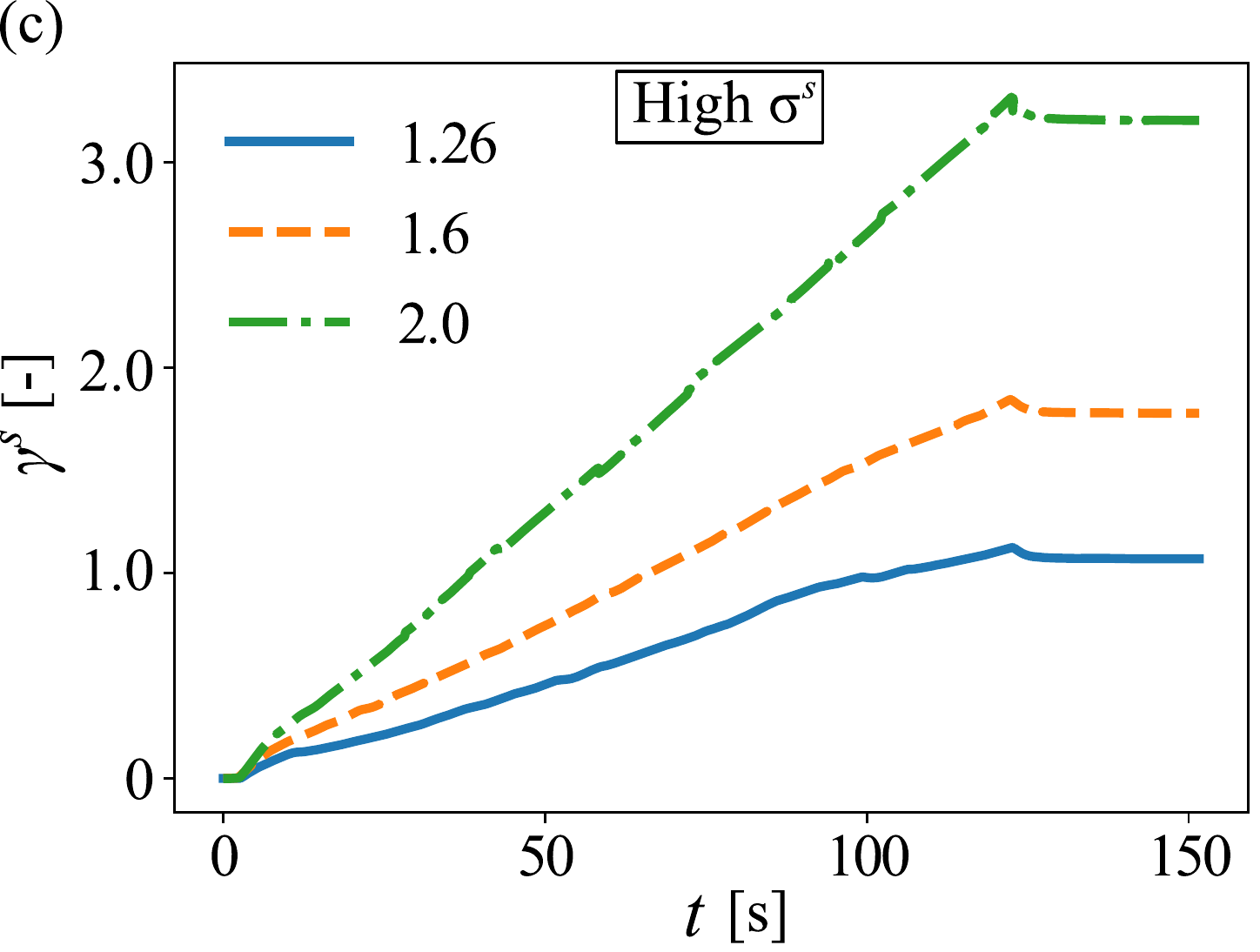}
    \caption{PMMA-laden interface at high surface coverage. (a)~Fluorescent confocal micrographs of PMMA particles (white) at an oil-water interface, $\phi= 56.7\%$. (b)~Corresponding strain vs time at low imposed rheometer rotation rates, $\omega$, see inset legend. $r_{\rm out}=\SI{6.6}{\milli\metre}$. $\omega$ applied from $t\gtrsim \SI{5}{\second}$ for \SI{120}{\second}, followed by \SI{30}{\second} further recording. (c)~Corresponding high imposed $\omega$ response.}
    \label{fig:highPhi}
\end{figure*}

We have seen that applying shear leads to considerable aggregation in this system. In order for aggregation to occur the shear force must exceed the maximum repulsive force between these particles. It has been observed that these particles have an interaction that can be described by a repulsive screened Coulomb potential~\cite{Muntz2020}. In order to overcome this repulsion we assume that they must overcome the maximum repulsive force. This maximum force can be found to be \SI{9.88D-13}{\newton}, where we have rescaled the parameters from Ref.~\onlinecite{Muntz2020}, as in this work we use larger particles. This force can then be converted to an interfacial stress by dividing by the particle diameter to give a critical aggregation stress of \SI{3.3D-7}{\pascal\metre}. 
Experimentally, we found that aggregation occurs starting at a stress of \SI{6.6D-7}{\pascal\metre}, which is in reasonable agreement. This agreement between experiment and prediction lends further confidence to the use of Eq.~\eqref{eq:surfaceStress} when calculating interfacial stress in our unique geometry. In order to aggregate, the applied stress must also overcome the steric repulsion, however the steric barrier is considerably smaller (for a similar particle) than the electrostatic barrier~\cite{Muntz2020, Mewis2000}. Once these particles come into close contact, attractive capillary forces and/or van der Waals forces are large enough such that this is irreversible.

\subsubsection{High surface coverage}\label{sec:high_surf}

Now, we investigate the same PMMA particles at higher surface fractions of 56.7\%, \partFig{fig:highPhi}{a}, where the particles assemble into a percolated aggregated structure. We observe markedly different behaviour, with elastic behaviour being evident from the strain vs time plots, \partFig{fig:highPhi}{b)--(c}. Focussing on the low stress behaviour, \partFig{fig:highPhi}{b}, an initial jump to a higher strain is observed, indicative of an elastic material. There is then some erratic motion in the direction of shear (\ie\ the strain is always positive), indicating that there is some frustrated motion and rearrangements of the interfacial structure~\cite{Weeks2002}. While the initial elastic response is difficult to measure precisely due to background noise in the flow, upon cessation of shear the interface clearly recoils, allowing the elastic strain to be readily measured~\cite{Imperiali2012}. This statement becomes even more apparent when looking at higher applied stresses, \partFig{fig:highPhi}{c}, as at these large stresses the flowing behaviour completely dominates the strain response and the initial elastic jump is barely visible in the data. However, once the shear has been stopped, the elastic recoil is clear. At the end of shear at low rotation rate we sometimes observe motion, \eg, \partFig{fig:highPhi}{b}~(solid line), perhaps due to thermal gradients or air flow --- note, however, that the shear rates are small.

\begin{figure}
    \centering
    \includegraphics[width = 0.95\columnwidth]{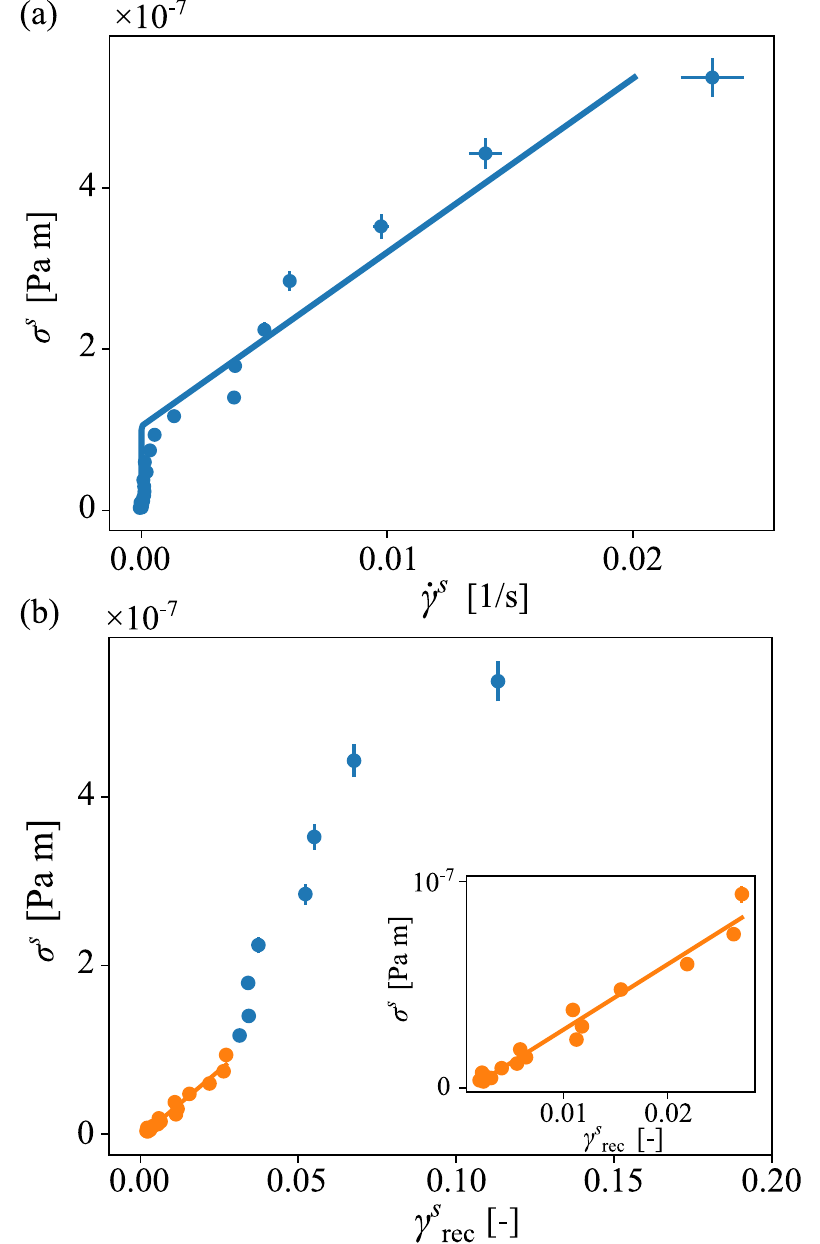}
    \caption{Rheological behaviour of an oil-water interface laden with PMMA particles at $\phi=56.7\%$. (a)~Viscous response to applied stress, \ie\ after initial elastic response, but before recoil. Points, data; line, fit to Bingham fluid model, Eq.~\eqref{eq:bingham}. (b)~Elastic response measured from recoil. Points, data; line, linear elastic fit to low strain, $\leq 0.03$ (orange and inset), $\elastic = \SI{3.16(16)D-6}{\pascal\metre}$.}
    \label{fig:190320_rheo}
\end{figure}

At higher $\phi$ we observe a more complex rheological response. By plotting the plastic response, stress vs shear rate, we can infer that the particle-laden interface behaves as a yield stress fluid, \partFig{fig:190320_rheo}{a}, as has been observed previously using the DWR geometry~\cite{VanHooghten2017}. By fitting a Bingham plastic model, Eq.~\eqref{eq:bingham}, we measure a yield stress of \SI{1.05(15)D-7}{\pascal\metre}. The effective interfacial viscosity (\SI{2.16(14)D-5}{\pascal\metre\second}) is, as expected, larger than that measured at the lower $\phi$, Fig.~\ref{fig:oilThickness}. We feel that this is an appropriate model, as the parameter which we are most interested in comparing to data in the published literature is the yield stress. The measured $\sigma^s_y$ is, however, an order of magnitude lower than the yield stress quoted in Ref.~\onlinecite{VanHooghten2017}. As close agreement between the techniques is found for a \pnip--laden interface, \partFig{fig:microgelRheology}{c}, this suggests that the different surface coverages may not be comparable, 56.7\% here vs 74\% in Ref.~\onlinecite{VanHooghten2017}. Moreover, there is a difference in PMMA stabilizer, poly(12-hydroxystearic acid) in Ref.~\onlinecite{VanHooghten2017}~and poly(lauryl methacrylate) here, though that only makes a small difference in contact angle and a relatively small difference in interaction potential \cite{Muntz2020}. 

When plotting the elastic response at higher $\phi$, stress vs recoverable strain [\partFig{fig:190320_rheo}{b}], the response is initially linear, with $\sigma^s$ and $\gamma^s_{\rm rec}$ proportional up to a strain of 0.03. We fit a linear dependence of the elastic strain response to the imposed shear stress in the low-strain regime [inset (orange) points]. This modelled Hookean behaviour gives us a shear modulus of \SI{3.16(16)D-6}{\pascal\metre}; this is a factor $10\times$ lower in stiffness compared to the \pnip\ interface, \partFig{fig:microgelRheology}{a}, and over a smaller linear region, leading to a far weaker interface ($\sim 30\times$). Previously reported measurements on a similar particle-laden interface~\cite{VanHooghten2017} found interfacial moduli of $\sim \SI{2D-6}{\pascal\metre}$, which is consistent with our measurements here being at a slightly lower surface coverage, similar to the difference in $\sigma^s_y$.

Finally, we look at the high-stress elastic response, which suggests a complex structural evolution. At $\gamma^s_{\rm rec}\gtrsim 0.03$, \partFig{fig:190320_rheo}{b} (blue), the recoverable strain initially remains unchanged with $\sigma^s$, no longer increasing linearly. The shift in response at $\sigma^s \!\approx \!\SI{1D-7}{\pascal\metre}$, correlates well with the yield stress, \partFig{fig:190320_rheo}{a}. So, just above $\sigma^s_y$ a fixed maximum $\gamma^s_{\rm rec}$ can be stored, similar to the \pnip\ interface, \partFig{fig:microgelRheology}{a}. However, as $\sigma^s$ increases further, $\gamma^s_{\rm rec}$ begins to increase more rapidly, \ie\ strain softening. Qualitatively, this aligns with a variety of literature results. For example, \citet{Reynaert2007}~measured the surface elastic modulus for polystyrene particle aggregates at a water-oil interface, which decreased with strain amplitude. \citet{Zhang2016}~used large amplitude oscillatory strain rheology and observed strain softening for weakly attractive silica nanoparticles at an air-aqueous interface. Finally, \citet{Orsi2012}~used an interfacial shear rheometer on gold nanoparticles at an air-water interface and also observed strain softening, which they attributed to breaking of weak bonds in a 2D gel. This suggests that the interface evolves above yielding, notably, in the stress range for aggregation at low $\phi$, \SI{3.3D-7}{\pascal\metre}, which should be $\phi$ independent. It is then possible that the strain softening is driven by aggregation, consistent with aggregation weakening interfaces at all but the highest $\phi$ (and lowering $\eta^s$, Sec.~\ref{sec:low_surfAgg}).

\subsection{Limits and comparison of techniques}

Our results, and the comparison to 1) DWR results using the same PNIPAM-SiO$_2$ system, Fig.~\ref{fig:microgelRheology}, and 2) similar colloidal particle laden interfaces using a DWR~\cite{VanHooghten2017} or magnetic rod~\cite{Reynaert2007}, suggest that our setup represents a useful addition to the field of interfacial rheology. In comparable situations, our results are highly consistent with results using conventional probes directly attached to the interface, \eg, the elastic modulus of the \pnip\ interface, or within expected variation due to differing methods or interface preparation, \ie\ the yield stress for both a \pnip\ or PMMA-particle laden system.  Crucially, our setup allows us to both measure viscosities at lower $\phi$ than have been previously observed using the DWR, and to probe static yielding at lower stresses. The lower stress limit can be estimated using Bo$^*\approx 10$, Eq.~\eqref{eqn:boussinesq_particle}, alongside an estimate of the minimum interfacial shear rate set by the imaging setup resolution. With a minimum resolvable strain of $\gamma_{\rm rec}^{s,\min} =0.002$, \partFig{fig:microgelRheology}{a}, over a \SI{120}{\second} experiment the minimum shear rate is $\dot\gamma^{s,\min}\sim \SI{2D-5}{\per\second}$, or a rotation rate $\omega_i^{\min} = \dot\gamma^s r_{\rm out} / (r_r-r_{\rm out}) \approx \SI{2.5D-5}{\radian\per\second}$. For pre-factors in Eq.~\eqref{eqn:boussinesq_particle}\,$\approx 1$, this sets $\omega^{\min} \approx 0.002$\,rpm. The minimum rotation rate then sets a lower stress limit, Eq.~\eqref{eq:surfaceStress}, $\sigma^{s,\min} = \mathcal{O}(\SI{e-8}{\pascal\metre})$, comparable to more sensitive interfacial techniques, \eg, a micro-needle~\cite{Tajuelo2015,Renggli2020}. To specifically probe low stresses, $\dot \gamma^{s,\min}$, and so $\sigma^{s,\min}$, could be further optimised by using a high magnification or longer imaging period, together with minimisation of noise (\eg, thermal gradients and vibrations). Most remarkably, our contactless technique retains the maximum stresses, \partFig{fig:microgelRheology}{c}, attainable for a DWR using a closed feedback loop~\cite{Renggli2020}, and hence has a dynamic range that spans the majority of interfacial shear rheometry methods. This wide range of measurable stress combined with \textit{in situ} determination of interfacial characteristics, either surface pressure via a Wilhelmy plate or surface fraction via imaging, opens up the contactless technique to multiple future applications.

\section{Conclusion}
In this work, we have developed a contactless method to perform interfacial shear rheology on liquid-liquid interfaces without an interfacial geometry. The shear is applied to the continuous phase using a rotational rheometer and indirectly deforming the interface and the surface response is measured via confocal microscopy, either of a fluorescent particle-laden interface or via tracer particles embedded in the interface, enabling the measurement of a broad range of interfaces formed from, \eg, proteins, polymers or molecular surfactants~\cite{Tcholakova2008}. While we use a confocal microscope and stress-controlled rheometer, the same results should be achievable using any fixed-rate motor and reflection or fluorescence microscopy with sufficient resolution and frame rate. This enhances the applicability of this method as only relatively common equipment is required.

The method has been verified using a \pnip--laden interface measured using both our novel contactless geometry and a conventional DWR method, with equal elastic moduli found and comparable yield stress values. Our contactless setup allows us to both measure interfacial viscosities at lower surface fractions than have been previously observed using the DWR, owing to the high sensitivity achieved by having no probe attached directly to the interface, while maintaining the ability to apply large interfacial stresses.  Additionally, we have linked the rheological behaviour to the structural behaviour of PMMA particle interfaces with different initial conditions. At low surface coverage, the interface behaves as a two-dimensional Newtonian fluid and is subject to aggregation above a certain shear threshold. At higher surface coverage the interface begins to behave as an elastic sheet with a measurable shear modulus, up to a yield stress where the interface begins to flow. In addition, our results suggest that both surface coverage and interfacial particle aggregation state affect the rheology of the interface, in line with results in the literature.

This work has focussed on the motion of the particles in the plane of the interface under steady shear. As the setup presented here does not have a probe attached to the liquid interface, the effect of the interface on how shear is propagated from the oil to the water phase can now be studied. This would facilitate observation of how the inside of an emulsion droplet is influenced by shear of the continuous phase, thereby greatly increasing the understanding and predictability of the flow behaviour of these systems, which are encountered ubiquitously in many formulation applications.

\section*{Acknowledgments}

IM acknowledges studentship funding from the EPSRC Centre for Doctoral Training in Condensed Matter Physics (CM-DTC, EP/L015110/1). SB acknowledges studentship funding from the EPSRC Centre for Doctoral
Training in Soft Matter and Functional Interfaces (SOFI-CDT, EP/L015536/1). MR acknowledges funding from the Marie Sklodowska-Curie Individual Fellowship
(Grant No.\ 101064381). The authors acknowledge R.~O'Neill, R.\ Van Hooghten, Damian Renggli and Jan Vermant for useful discussions, A. Garrie for the PTFE cup and aluminium ring, J.~Royer for assistance with the rheoimaging setup, and M.~Hermes for the velocimetry C code. For the purpose of open access, the authors have applied a Creative Commons Attribution (CC BY) licence to any Author Accepted Manuscript version arising from this submission.
\section*{Author declarations}
\subsection*{Conflicts of Interest}
The authors declare that they have no conflicts of interest.
\subsection*{Data Availability}
The data that support the findings of this study are openly available in Edinburgh DataShare at  \href{https://doi.org/10.7488/ds/3759}{https://doi.org/10.7488/ds/3759}.
\appendix*
\section{Surface Coverage Measurements\label{app:surfaceFraction}}
As small differences in surface fraction can cause significant differences in rheological response, and surface fractions for non-close packed PMMA monolayers formed by sedimentation are challenging to reproduce, the surface coverage $\phi$ is determined for each interface prepared. First, microscopy images are analysed, the magnification level being a balance between individual particle resolution, which improves upon magnification, and statistics of particle counting, which decreases upon magnification. Examples are shown in \partFig{fig:oilThickness}{b}. Surface fractions are quoted in terms of a pixel counting method, as described in the main text. To determine how accurately surface coverage can be defined, we compare the pixel fraction method with a particle counting method. Here, we can count the number of particles and use knowledge of particle size and image size to calculate surface fraction. For the sample in \partFig{fig:lowPhi}{a}, measurements of surface fraction via the pixel fraction and the particle counting methods give respectively 31.0\% and 23.0\% while for the sample in \partFig{fig:highPhi}{a} these give respectively 56.7\% and 46.1\%.

Note the significant difference in these measurements, highlighting the challenge in defining the surface coverage. With perfect particle resolution, the particle counting method should yield the correct answer \textit{for that particular region of the interface}, however, perfect particle resolution is rarely achieved (especially at high $\phi$), for instance because an aggregate could be mistaken for one particle. The pixel fraction method is also flawed in that it assumes a direct match between area of emitted light with area of the particle. This however is not true due to the point spread function of the imaging setup, a question over whether the particles are exactly in the focal plane, and the brightness of the fluorophore itself, among other considerations. Comparing these images with other, similar images taken using the same imaging setup allows us to have an estimate for the surface fraction and certainly allows us to observe trends in flow behaviour as surface fraction changes.

It is also worth noting in \partFig{fig:highPhi}{a} there appears to be two types of particle with different intensity levels. There are a few possible reasons: firstly, dispersity in particle size, fluorophore intensity, or contact angle~\cite{Isa2011} may lead to this effect. This aligns with some particles in \partFig{fig:lowPhi}{a} appearing smaller but with less appreciable change in intensity, presumably as the excitation signal is at saturation in these imaging conditions. Secondly, the particles, while in close contact, may be pushed out of the surface leading to variations in their vertical $z$-position. This, however, is unlikely as the $z$ resolution is $\gg$ particle diameter~\cite{Cole2011},
\begin{equation}
    \delta z = \frac{0.88\lambda_{\rm exc}}{1-\sqrt{(n^2-{\rm NA}^2)}} = \SI{9.3}{\micro\metre},
\end{equation}
for excitation wavelength $\lambda_{\rm exc} = \SI{488}{\nano\metre}$, dry objective refractive index $n = 1$, and numerical aperture ${\rm NA} = 0.3$. Finally, aggregates might cause a (local) curvature in the liquid-liquid interface, which could lead to some particles remaining interfacial but moving relative to the imaging plane, which would result in particles appearing smaller and/or less bright.

These possible artefacts highlight the challenge in precisely determining the surface fraction from images. However, the pixel counting method gives a consistent result that allows a systematic comparison with increasing surface coverage of particles, Sec.~\ref{sec:PMMAresults}, and a reasonable comparison with the literature results of \eg, Ref.~\onlinecite{VanHooghten2017}.

\section*{References}

%

\end{document}